\documentclass[aps,prx,twocolumn,amsmath,amssymb,superscriptaddress]{revtex4}

\usepackage[colorlinks,citecolor=blue,linkcolor=Fuchsia]{hyperref}
\usepackage[usenames,dvipsnames,svgnames]{xcolor}

\usepackage{graphicx}
\usepackage{Jonasmacros}
\usepackage{bm}
\usepackage{mathptmx}
\usepackage{txfonts}
\usepackage{units}
\renewcommand{\vec}[1]{\boldsymbol{#1}}

\begin{document}  
\date{\today}  
\title{Microscopic Theory for Coupled Atomistic Magnetization and Lattice Dynamics}

\author{J. Fransson} 
\email{Jonas.Fransson@physics.uu.se} 
\affiliation{Department of Physics and Astronomy, Box 516, SE-751 20, Uppsala University, Uppsala, Sweden} 

\author{D. Thonig}
\affiliation{Department of Physics and Astronomy, Box 516, SE-751 20, Uppsala University, Uppsala, Sweden} 

\author{P. F. Bessarab}
\affiliation{Science Institute of the University of Iceland, 107 Reykjavik, Iceland}
\affiliation{ITMO University, 197101 St. Petersburg, Russia}

\author{S. Bhattacharjee}
\affiliation{Indo-Korea Science and Technology Center (IKST), Bangalore, India}

\author{J. Hellsvik}
\affiliation{Nordita, Roslagstullsbacken 23, SE-106 91 Stockholm, Sweden}
\affiliation{Department of Physics, KTH Royal Institute of Technology, SE-106 91 Stockholm, Sweden}

\author{L. Nordstr\"om} 
\affiliation{Department of Physics and Astronomy, Box 516, SE-751 20, Uppsala University, Uppsala, Sweden} 
 
\begin{abstract}
A coupled atomistic spin and lattice dynamics approach is developed which merges the dynamics of these two degrees of freedom into a single set of coupled equations of motion. The underlying microscopic model comprises local exchange interactions between the electron spin and magnetic moment and the local couplings between the electronic charge and lattice displacements. An effective action for the spin and lattice variables is constructed in which the interactions among the spin and lattice components are determined by the underlying electronic structure. In this way, expressions are obtained for the electronically mediated couplings between the spin and lattice degrees of freedom, besides the well known inter-atomic force constants and spin-spin interactions. These former susceptibilities provide an atomistic ab initio description for the coupled spin and lattice dynamics. It is important to notice that this theory is strictly bilinear in the spin and lattice variables and provides a minimal model for the coupled dynamics of these subsystems and that the two subsystems are treated on the same footing. Questions concerning time-reversal and inversion symmetry are rigorously addressed and it is shown how these aspects are absorbed in the tensor structure of the interaction fields. By means of these results regarding the spin-lattice coupling, simple explanations of ionic dimerization in double anti-ferromagnetic materials, as well as, charge density waves induced by a non-uniform spin structure are given. In the final parts, a set of coupled equations of motion for the combined spin and lattice dynamics are constructed, which subsequently can be reduced to a form which is analogous to the Landau-Lifshitz-Gilbert equations for spin dynamics and damped driven mechanical oscillator for the ionic motion. It is important to notice, however,  that these equations comprise contributions that couple these descriptions into one unified formulation. Finally, Kubo-like expressions for the discussed exchanges in terms of integrals over the electronic structure and, moreover, analogous expressions for the damping within and between the subsystems are provided. The proposed formalism and new types of couplings enables a step forward in the microscopic first principles modeling of coupled spin and lattice quantities in a consistent format.
\end{abstract}
 
\maketitle

\section{Introduction}
The understanding of how spin and lattice degrees of freedom interact is of fundamental importance \cite{Jensen1991rem,Dong2015}. Recently, strong evidence was found for the existence of a significant coupling between magnons and phonons for instance in bcc Fe \cite{Koermann2014,Perera2017} and the ferromagnetic semiconductor EuO \cite{Pradip2016}. Spin-lattice coupling is central for seemingly disparate phenomena such as the mechanical generation of spin currents by spin-rotation coupling \cite{Matsuo2013}, the spin-Seebeck effect \cite{Jaworski2011,Wilken2016PhD}, and the driving of magnetic bubbles with phonons \cite{Ogawa2015}. In the field of multiferroic spin-lattice coupling is a central mechanism for the coupling of (anti)ferromagnetic and (anti)ferroelectric order parameters (magnetoelectric effect) \cite{Kimura2003,Pimenov2006,Sushkov2008,Dong2015}.
Spin-lattice coupling also occur in ferroelastic and ferromagnetic materials (magnetoelastic effect) \cite{McCorkle1923}. 
There is also a growing interest in including effects from mechanical degrees of freedom into theoretical models for ultrafast magnetization dynamics \cite{Kimel2009,Ostler2012,Johnson2012}, since rapid ionic motion has shown to cause non-trivial temporal fluctuations of the magnetic properties \cite{Wall2009,Kim2012,Nova2016,Juraschek2017b}.

Magnetization dynamics is conventionally understood in terms of the phenomenological Landau-Lifshitz-Gilbert \cite{Landau35,Gilbert2004} approach. A seminal step towards a formulation of atomistic magnetization dynamics from first principles was taken by Antropov~\emph{et al.}~\cite{Antropov1996} who started out from time-dependent density functional theory and the Kohn-Sham equation and considered also simultaneous spin and molecular dynamics, however, incorporating energy dissipation and finite temperature phenomenologically. The equation of motion for local spin magnetic moments in the adiabatic limit have also been worked out in Refs.~\cite{Niu1999,Qian2002}. Effects of non-locality in space and time were captured in the formalism communicated in \cite{Bhattacharjee2012}, including a complete basic principle derivation of the atomistic magnetization dynamics equations of motion. 

Recently, great effort has been devoted to improve the Landau-Lifshitz-Gilbert approach by calculating the damping tensor directly from the electronic structure \cite{Ebert2011,Thonig2014,Gilmore2007}. The addition of other contributions, as for instance moment of inertia \cite{Ciornei2011,Bhattacharjee2012,Thonig2017} observed in Refs. \cite{Fahnle2011,Fahnle2013,Ciornei2011}, allows for dynamics on shorter time scales. The basic principles of the moment of inertia contributions to atomistic magnetization dynamics were derived from a Lagrangian formulation \cite{Wegrowe2012}. In the adiabatic limit, the lattice degrees of freedom follow Newton dynamics \cite{Forshaw2009dar} and can be derived from the effective action of the system \cite{Nolting2016tp2}. Hence, the uncoupled dynamics of spin and lattice is well understood \cite{Eriksson2017asd,Nolting2016tp2}.

There have been in the last years been several simulations with a combined Landau-Lifshitz-Gilbert and lattice dynamics approach \cite{Ma2008,Ma2012}. They are based on an atomistic spin model with position dependent exchange parameters which for instance lead to a spin ordering dependent effective lattice dynamics equations of motion. 
This spin and lattice coupling then enter through the Taylor expansion of the magnetic exchange interactions in terms of ionic displacements around the equilibrium positions.

To put spin and lattice degrees of freedom on the same footing, however, bilinear order of spin-lattice coupling is required that seems forbidden from the naive argument of breaking the time reversal symmetry in the total energy. Thus, the question remains about the lowest order in spin-lattice coupling, conserving Newtons third law.

\begin{figure}
\centering
 \includegraphics[width = 0.9\columnwidth]{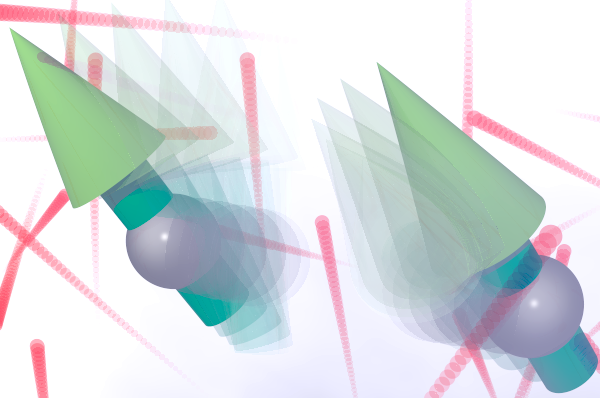}
\caption{(Color online) Schematic figure of two atoms (gray balls) with a magnetic moment (green arrows) and lattice vibrations (fading gray and transparent balls) in a cloud of electrons (small red balls and foggy environment) }
\label{fig:DAFM}
\end{figure}

We notice that in the past there have been several considerations of coupling magnetic and elastic degrees of freedom, see for instance Refs. \cite{PhysRev.110.836,JETP.35.157,JMathPhys.5.1298}. A bi-linear magneto-elastic coupling, which has some similarities to the coupling derived in this paper, has also been considered previously \cite{PhysRev.110.836}. However, all these discussions were based on hydrodynamics approaches aiming towards phenomenological descriptions of the macroscopic continuum and mechanisms for coupling between magnetic and elastic properties of solids. Such accounts are accreditable  only in the long wave length limit. We find that there is an apparent lack in the literature of systematic descriptions addressing the quantum mechanical nature of metals which is responsible for the effective couplings between the degrees of freedom represented by the spins and lattice at an atomistic length scale.

The purpose of this Paper is to derive from first principles a theoretical framework for coupled atomistic magnetization and lattice dynamics. In order to treat magnetic and mechanical degrees of freedom on the same footing, our starting point is to formulate the action of the system. From this action we derive, to leading order, bilinear couplings between spins and mechanical displacements, couplings which are of three different types, namely spin-spin, displacement-displacement and the novel bilinear spin-displacement coupling. Furthermore we obtain the coupled equations of motion for the mechanical displacement $\{\bf Q_i\}$ and velocity $\{\bf V_i\}$, and the magnetization $\{\bf M_i\}$ dynamics, thus providing a natural extension of harmonic lattice dynamics, on the one hand, and the LLG description of the magnetization dynamics of bilinear spin Hamiltonians, on the other. The framework is applicable to general out-of-equilibrium conditions and includes also retardation mechanisms.

In general terms we address the question whether the electrons in a metal that, on the one hand are influenced by the ionic vibrations, or, phonons, through the electron-phonon coupling and, on the other hand, couple to magnetic moments via exchange, thereby mediates an interaction between the ionic vibrations and the magnetic moments. With this question in mind, we derive a general minimal model for the magnetic and mechanical degrees of freedom where the interactions between the entities are mediated by the underlying electronic structure. We show that the effective model comprises both the well known bi-linear magnetic indirect exchange interaction as well as the electronic contribution to the interatomic force constant. However, the derivation also shows the existence of a bi-linear coupling between the magnetic and mechanical entities. The present paper is essentially focused on this derivation and the properties of the bi-linear spin-lattice coupling from microscopic theory.

The Paper is organized as follows. In Sec. \ref{sec-derivation} we derive a complete and generalized spin-lattice model. The related bilinear spin-lattice Hamiltonian and its inherent symmetries of the are discussed in Secs. \ref{app-bilinear} and \ref{sec-symmetries} and a few numerical examples are studied in Sec.~\ref{sec-examples}. In Sec. \ref{sec-taylor} we make a brief comparison to expanding the exchange parameters as function of spatial coordinates. The dynamics of coupled spin-lattice reservoirs are evaluated in Sec.\ref{sec-eqm} and the paper is summarized in Sec. \ref{sec-summary}. Further details are given in the appendix.

\section{Derivation of effective spin-lattice model}
\label{sec-derivation}

\subsection{Effective action}
\label{ssec-action}
The effective action for the coupled spin-lattice system is constructed and analysed. In absence of any \emph{ad-hoc} coupling between the spin and lattice subsystems, we address the full microscopic model of the material through the partition function
\begin{align}
\calZ=&
	e^{i\calS}
	,
\end{align}
where the total action $\calS$ is given by
\begin{align}
\calS=&
	\calS_0
	+
	\calS_\text{latt}
	+
	\calS_\text{WZWN}
	+
	\calS_B
	+
	\calS_E
	+
	\oint
	\Bigl(
		\Hamil_M
		+
		\Hamil_\text{ep}
	\Bigr)
	dt
	.
\end{align}
Instead of expressing all components in mathematical terms here, we discuss the physics involved in each contribution and refer to Appendix \ref{app-derivation} for details.

Accordingly, $\calS_0$ accounts for the part of the electronic structure that does not directly relate to the localized spin moments $\bfM$ and lattice displacements $\bfQ$, whereas $\calS_\text{latt}$ provides the analogous components for the unperturbed lattice vibrations. As for the latter, we shall not make any assumptions about the model for the lattice dynamics but notice that the mechanism for the coupling between the spin and lattice subsystems does not depend on the specifics of the lattice model. Accordingly, the intrinsic lattice vibrations can be treated to any order of accuracy. Furthermore, the Wess-Zumino-Witten-Novikov component $\calS_{WZWN}$ accounts for the Berry phase accumulated by the spin motion, whereas $\calS_B$ and $\calS_E$ comprise the coupling to the external magnetic and electric fields, respectively. Finally, the Hamiltonian $\Hamil_M$ describes the Kondo coupling between the itinerant electron spin $\bfs_s\equiv\psi^\dagger\bfsigma\psi/2$ and the localized spin moment $\bfM$ while $\Hamil_\text{ep}$ provides the coupling between the electronic charge $n=\bfs_c\equiv\psi^\dagger\sigma_0\psi$ and the lattice displacements $\bfQ$, or in other words, the electron-phonon coupling. Here, also $\psi=(\psi_\up\ \psi_\down)^T$ is the electron spinor, $\sigma_0$ is the $2\times2$ identity, and $\bfsigma$ is the vector of Pauli matrices.

Given the above structure we can address both equilibrium \emph{and} non-equilibrium problems by defining the quantities appropriately either to a well defined ground state in the former case or by expanding the time integration to the Keldysh contour and relate the physics to some initial state defined in the far past in the latter. We, therefore, keep the derivation as general as possible and choose the latter approach as the generic one. Despite the additional complexity this route entails, it is justified since the equilibrium physics can always be retained from the non-equilibrium description.

\subsection{Dynamical bi-linear couplings}
\label{ssec-bilinear}
We obtain the effective action $\calS_\text{MQ}$ for the coupled magnetization and lattice dynamics through a second order cumulant expansion of the partition function subsequently followed by tracing over the electronic degrees of freedom. The resulting model can be written
\begin{align}
\calS_\text{MQ}=&
	-\frac{1}{2}\int
	\Bigl(
			\bfQ(x)\cdot
			[
				\calT_{cc}(x,x')\cdot\bfQ(x')
				+\calT_{cs}(x,x')\cdot\bfM(x')
			]
\nonumber\\&
			+\bfM(x)\cdot
			[
				\calT_{sc}(x,x')\cdot\bfQ(x') 
				+\calT_{ss}(x,x')\cdot\bfM(x')
			]
	\Bigr)
	dxdx',
\label{eq-effS} 
\end{align}
where we have introduced the notation $x=(\bfr,t)$ and defined the interaction tensor
\begin{subequations}
\begin{align} 
\calT_{pq}(x,x')=& 
	\int
		\Xi_p(\bfr,\bfrho)
		\bfK_{pq}(y,y')
		\Xi_q(\bfrho',\bfr')
	d\bfrho
	d\bfrho'
	,
\label{eq-Dpq}
\\
\bfK_{pq}(y,y')=&
	\eqgr{\bfs_p(y)}{\bfs_q(y')}
	,\ 
	 y=(\bfrho,t)
	 ,\
	 p,q=c,s
	 .
\label{eq-Kpq}
\end{align}
\end{subequations}
Here, the parameters $\Xi_c(\bfr,\bfr')$ and $\Xi_s(\bfr,\bfr')$ define the electron-phonon and Kondo coupling, respectively, and we have adopted the notation where the subscript $c$ refers to charge and $s$ to spin.

The effective model given in Eq. (\ref{eq-effS}) can be reduced to an analogous lattice model, the bi-linear Hamiltonian which can be written as
\begin{align}
\Hamil_\text{MQ}=&
	-\frac{1}{2}
	\sum_{ij}
	\Bigl(
			\bfQ_i\cdot
			[
				T^{cc}_{ij}\cdot\bfQ_j
				+
				T^{cs}_{ij}\cdot\bfM_j
			]
\nonumber\\&
			+\bfM_i\cdot
			[
				T^{sc}_{ij}\cdot\bfQ_j
				+
				T^{ss}_{ij}\cdot\bfM_j
			]
	\Bigr)
	,
\label{eq-effHdiscrete}
\end{align}
where we denote the magnetic moment centered at the atomic position $i$ as $\bfM_i$ and the local atomic displacements as $\bfQ_i$, where the here instantaneous lattice interactions tensors are denoted as $T_{ij}^{pq}$.

The effective model presented here, demonstrates the presence of a bilinear coupling $\calT_{sc/cs}$ between the spin and lattice subsystems. It also indicates that this coupling is mediated by the background electronic structure of the material in analogous forms as the spin-spin interactions $\calT_{ss}$ as well as the lattice-lattice coupling, or, the electronic contribution to the interatomic force constant $\calT_{cc}$. Although this is not surprising, given the set-up of the system, it is nonetheless an important observation since it demonstrates the lowest order of indirect exchange interaction between the spin and lattice subsystems and, since it is generated by the same interaction field as the spin-spin and lattice-lattice couplings, it is expected to have a non-trivial impact on certain classes of materials. It is therefore of utter importance to derive expressions for the spin-lattice couplings in order to both compare to the spin-spin/lattice-lattice interactions but also to enable a deeper analysis and understanding of which condition that have to be fulfilled to create finite spin-lattice couplings.

For the sake of argument we, therefore, decouple the propagator $\bfK_{pq}$ into a product of two single electron Green functions $\bfG$, see Appendix \ref{app-derivation}, which are defined by the background electronic structure, given by the Hamiltonian $\Hamil_0$. It is then straight forward to derive
\begin{align}
\bfK_{pq}(x,x')=&
	(-i)
	{\rm sp}
	\bfsigma_p
	\bfG(x,x')
	\bfsigma_q
	\bfG(x',x)
	/2^{\delta_{ps}+\delta_{qs}}
	,
\end{align}
where ${\rm sp}$ denotes the trace over spin space and where $\bfsigma_c=\sigma_0$ and $\bfsigma_s=\bfsigma$.

Next, since the Hamiltonian can be partitioned into charge and spin components according to $\Hamil_0=\Hamil_0^{(0)}\sigma_0+\Hamil_0^{(1)}\cdot\bfsigma$, the analogous partitioning can be made for the Green function $\bfG$ in terms of charge and spin components $G_0$ and $\bfG_1$, respectively. Thus, we can write $\bfG=G_0\sigma_0+\bfG_1\cdot\bfsigma$. Using these two observations, one immediately obtains
\begin{align}
{\rm sp}&
	\bfsigma_p\bfG\bfsigma_q\bfG
	=
	{\rm sp}
	\bfsigma_p
	\Bigl(
		G_0+\bfG_1\cdot\bfsigma
	\Bigr)
	\bfsigma_q
	\Bigl(
		G_0+\bfG_1\cdot\bfsigma
	\Bigr)
	.
\label{eq-spG}
\end{align}
By tracing over the spin degrees of freedom, the nature of the lattice-lattice, spin-lattice, and spin-spin interactions can be further analyzed in terms of the Green function components that constitute the expressions.


As one of the purposes with this paper is to construct a coherent formalism for the coupled spin and lattice dynamics, we present the results for all three types of couplings. The details of the derivations can be found in Sec.~\ref{app-bilinear}.

\subsection{Lattice-lattice coupling}
\label{ssec-pp}
Setting $p=q=c$ in Eq. (\ref{eq-spG}), the interaction tensor describes the electronic contribution to the interatomic force constant $\Phi(x,x')\equiv\calT_{cc}(x,x')$. Putting the coupling $\Xi_c(\bfr,\bfr')=\bflambda(\bfr,\bfr')$, where $\bflambda(\bfr,\bfr')$ is the local electron-phonon coupling, see Sec. \ref{aapp-microscopic} for more details, the interatomic force constant acquires the form
\begin{align}
	\calT_{cc}(x,x')=&
	(-i)2
	\int
		\bflambda(\bfr,\bfrho)
		\Bigl(
			G_0(y,y')G_0(y',y)
\nonumber\\&
			+
			\bfG_1(y,y')\cdot\bfG_1(y',y)
		\Bigr)
		\bflambda(\bfrho',\bfr')
	d\bfrho
	d\bfrho'
	.\label{eq:dynspinlatt}
\end{align}
The interatomic force constant is therefore a direct measure of the total electronic structure to which the lattice vibrations are coupled. Moreover, although there is no directionality induced by the spin texture ($\bfG_1$) in the electronic structure, its makes an important contribution to the overall interaction strength.
%
It can also be seen that the tensorial structure of the interactions is governed by the structure factor $\bflambda$ of the electron-phonon coupling, as the dyad $\bflambda\bflambda=\lambda_i\lambda_j\hat{\bf i}\hat{\bf j}$.

\subsection{Spin-spin coupling}
\label{ssec-ss}
In case of the spin-spin coupling we put $p=q=s$ in Eq. (\ref{eq-spG}), for which we obtain
\begin{align}
\bfM(x)\cdot\calT_{ss}(x,x')\cdot\bfM(x')=&
	\calJ(x,x')\bfM(x)\cdot\bfM(x')
\nonumber\\&
	+
	\calD(x,x')\cdot
	\Bigl(
		\bfM(x)\times\bfM(x')
	\Bigr)
\nonumber\\&
	+
	\bfM(x)\cdot\calI(x,x')\cdot\bfM(x')
	,
\end{align}
where the three contributions represent the isotropic Heisenberg, and the anisotropic Dzyaloshinskii-Moriya and Ising interactions, respectively. The order of these contributions is natural since they are the rank 0, 1, and 2 tensors emerging from the general rank 2 tensor $\calT_{ss}$. It should also be noticed that the first ($\calD$) and second ($\calI$) rank tensors represent the anti-symmetric and symmetric contributions to the exchange \cite{Udvardi2003}. Similarly as for the interatomic force constant $\Phi$, we can write
\begin{subequations}
\label{eq-genDss}
\begin{align}
\calJ(x,x')=&
	-\frac{i}{2}
	\int
		\nu(\bfr,\bfrho)
		\Bigl(
			G_0(y,y')G_0(y',y)
\nonumber\\&
			-
			\bfG_1(y,y')\cdot\bfG_1(y',y)
		\Bigr)
		\nu(\bfrho',\bfr')
	d\bfrho
	d\bfrho'
	,
\label{eq-genJ}
\\
\calD(x,x')=&
	\frac{1}{2}
	\int
		\nu(\bfr,\bfrho)
		\Bigl(
			G_0(y,y')\bfG_1(y',y)
\nonumber\\&
			-
			\bfG_1(y,y')G_0(y',y)
		\Bigr)
		\nu(\bfrho',\bfr')
	d\bfrho
	d\bfrho'
	,
\label{eq-genDM}
\\
\calI(x,x')=&
	-\frac{i}{2}
	\int
		\nu(\bfr,\bfrho)
		\Bigl(
			\bfG_1(y,y')\bfG_1(y',y)
\nonumber\\&
			+
			[\bfG_1(y,y')\bfG_1(y',y)]^T
		\Bigr)
		\nu(\bfrho',\bfr')
	d\bfrho
	d\bfrho'
	.
\label{eq-genI}
\end{align}\label{eq-dyn-ss}
\end{subequations}
These expressions clearly illustrate that the Heisenberg interaction is finite independently on whether the background electronic structure has a spin texture ($\bfG_1$) or not, whereas both the Dzyaloshinskii-Moriya and Ising interactions are finite only in materials with non-vanishing spin texture, that is, either a simple spin-polarization and/or a non-collinear magnetic structure. Here, $\Xi_s(\bfr,\bfr')=\nu(\bfr,\bfr')$, where $\nu(\bfr,\bfr')$ is the direct exchange contribution from the Coulomb integral, see Sec. \ref{aapp-microscopic} for more details. Eq.~\eqref{eq-genDM} is in agreement with the expression for Dzyaloshinskii-Moriya in Ref.~\cite{Mankovsky2017pp}.

The anti-symmetric properties of $\calD$ is also clearly illustrated by Eq. (\ref{eq-genDM}), since interchanging the spatial coordinates is accompanied by a sign change, that is, $\calD(\bfr,\bfr';t,t')=-\calD(\bfr',\bfr;t,t')$, which signifies the odd property under spatial reversal. While this property can be obtained, e.g., in structures with finite spin-orbit  coupling, it can also be finite in general spatially inhomogeneous structures with non-collinear magnetic texture \cite{Fransson2010}. These observations accordingly suggest that a Dzyaloshinskii-Moriya interaction can be engineered in hetero-structures and tunnel junctions \cite{Zhu2006,Fransson2008,Fransson2008b,Fransson2014}.

The Ising interaction, finally, is the symmetric part of the tensor and it is finite in materials with a finite spin-polarization in the background electronic structure and both for a trivial or non-trivial spin texture \cite{Imamura2004,Fransson2010,Lounis2008,Fransson2014}. Hence, a simple spin-polarization along the $\hat{\bf z}$-axis generates a finite $I_{zz}\hat{\bf z}\hat{\bf z}$ while all other components of $\calI$ vanish. The contribution to the spin model then is $I_{zz}(x,x')S_z(x)S_z(x')$, which is the usual Ising model for collinear spins and the reason for calling it the Ising interaction.

\subsection{Spin-lattice coupling}
\label{ssec-sp}
Here, we finally discuss the new type of bi-linear interaction that we propose in this paper, namely, the spin-lattice coupling. Here, we set either $p=c$, $q=s$ in Eq. (\ref{eq-spG}), or the other way around, and for completeness we write both forms given by
\begin{subequations}
\begin{align}
\calT_{cs}(x,x')=&
	(-i)
	\int
		\bflambda(\bfr,\bfrho)
		\Bigl(
			G_0(y,y')\bfG_1(y',y)
			+
			\bfG_1(y,y')G_0(y',y)
\nonumber\\&
			-
			i
			\bfG_1(y,y')\times\bfG_1(y',y)
		\Bigr)
		\nu(\bfrho',\bfr')
	d\bfrho
	d\bfrho'
	,
\label{eq-genDcs}
\\
\calT_{sc}(x,x')=&
	(-i)
	\int
		\nu(\bfr,\bfrho)
		\Bigl(
			G_0(y,y')\bfG_1(y',y)
			+
			\bfG_1(y,y')G_0(y',y)
\nonumber\\&
			+
			i
			\bfG_1(y,y')\times\bfG_1(y',y)
		\Bigr)
		\bflambda(\bfrho',\bfr')
	d\bfrho
	d\bfrho'
	.
\label{eq-genDsc}
\end{align}
\end{subequations}
Here, we first notice that the electronically mediated spin-lattice coupling exists only in materials with either broken time-reversal symmetry and/or broken inversion symmetry, which is manifest in the explicit dependence on $\bfG_1$. Secondly, it can be noticed that the first two contributions to $\calT_{cs}$ and $\calT_{sc}$ are equal while the third contribution have opposite signs to one another. This structure reflects the composition of the tensor into one inversion symmetric and one inversion anti-symmetric component.

It is, moreover, interesting that the inversion symmetric component has an anti-symmetric time-reversal symmetry while the opposite observation can be made for the inversion anti-symmetric component. These properties are necessary in order to maintain the even properties of the effective spin model under both inversion and time-reversal symmetry operations. Hence, the result is that we can interchange the coordinates in, say, the contribution $\bfQ(x)\cdot\calT_{cs}(x,x')\cdot\bfM(x')$ in Eq. (\ref{eq-effS}), and from the conclusions in this section it follows that this contribution equals the other spin-lattice contribution, such that it is only necessary to write $2\bfQ(x)\cdot\calT_{cs}(x,x')\cdot\bfM(x')$ in the effective action. Therefore, the opposite signs of the inversion anti-symmetric contributions to $\calT_{cs}$ and $\calT_{sc}$ ensures that the correct symmetries are maintained for the spin-lattice model.

Further aspects regarding the symmetry properties will be discussed in Sec.~\ref{sec-symmetries}.

\section{Static Bi-linear Couplings}
\label{app-bilinear}

The properties of the bilinear couplings $\calT_{pq}(x,x')$ that we have introduced can be further analyzed in the static limit ($\omega\rightarrow 0$), that is, $\calT_{pq}^r(\bfr,\bfr')\equiv\lim_{\omega\rightarrow0}\calT_{pq}^r(\bfr,\bfr';\omega)=\lim_{\omega\rightarrow0}\int\calT_{pq}(\bfr,\bfr;t-t')e^{i\omega(t-t')}dt'$. Then, the general static  interaction tensor can be written as
\begin{subequations}
\begin{align}
\calT_{pq}^r(\bfr,\bfr')=&
	\int
		\Xi_p(\bfr,\bfrho)
		\bfK^r_{pq}(\bfrho,\bfrho')
		\Xi_q(\bfrho',\bfr')
	d\bfrho
	d\bfrho'
\label{eq-Tpq}
\\
\bfK^r_{pq}(\bfr,\bfr')=&
	-\frac{2}{2^{\delta_{ps}+\delta_{qs}}\pi}
	{\rm sp}
	\im
	\int
		f(\dote{})
			\bfsigma_p
			\bfG^r(\bfr,\bfr')
			\bfsigma_q
			\bfG^r(\bfr',\bfr)
	d\dote{}
	,
\label{eq-Kpq}
\end{align}
\end{subequations}
where the notation $\bfG^r(\bfr,\bfr')\equiv\bfG^r(\bfr,\bfr';\dote{})$.
This results is obtained by noticing that in equilibrium, the retarded susceptibility $\bfK_{pq}^r$ can be written as
\begin{align}
\bfK^r_{pq}(\bfr,\bfr';\omega)=&
	\frac{1}{2^{\delta_{ps}+\delta_{qs}}}
	{\rm sp}
	\int
		\frac{f(\dote{})-f(\dote{}')}{\omega-\dote{}+\dote{}'+i\delta}
\nonumber\\&\times
		\bfsigma_p
		\Bigl(-2\im\bfG^r(\bfr,\bfr')\Bigr)
		\bfsigma_q
		\Bigl(-2\im\bfG^r(\bfr',\bfr)\Bigr)
	\frac{d\dote{}}{2\pi}
	\frac{d\dote{}'}{2\pi}
	.
\end{align}
Then, by application of the Kramers-Kr\"onig relations, the result in Eq. (\ref{eq-Kpq}) follows.

A tool that is convenient to introduce for further discussion is a partitioning of the single electron Green functions according to 
\begin{align}
\bfG=G^0\sigma^0+\bfG_1\cdot\bfsigma=(G^{00}+G^{01})\sigma^0+(\bfG^{10}+\bfG^{11})\cdot\bfsigma\,.\label{eq-2index}
\end{align}
 Here, the first superscript 0 (1) refers to charge (spin) quantities, whereas the second superscript denotes whether the Green function { is even, 0, or odd, 1, under space reversal or equivalently change of direction $\bfr \rightleftarrows\bfr'$. Then the even Green functions, $G^{00}$ and $\bfG^{10}$, carry information about the charge and spin densities, respectively, while} the odd Green functions, $G^{01}$ and $\bfG^{11}$, are related to possible charge and spin currents, respectively, that may occur in the system. This means that only these Green functions may be finite under the current operator $\sim\nabla_\bfr-\nabla_{\bfr'}$ in the limit $\bfr'\rightarrow\bfr$. In summary, these four Green functions can be characterized in terms of being even and/or odd {under spin and space reversion} 
  as is illustrated in Table \ref{tab-GFs}. {In this Table we also summarize how they behave under time reversal. Under such an operation not only the spin but also the currents change sign, so  $G^{00}$ and $\bfG^{11}$ are invariant under time reversal while $G^{01}$ and $\bfG^{10}$ change sign.}
  
{An advantage with this formalism is that it becomes straight forward to study the effect of spin-orbit (spin-orbit) coupling. This is because for topologically trivial magnetic systems in equilibrium, the odd space reversal Green functions are odd in the spin orbit coupling while the even functions are even. Hence, in the absence of spin-orbit coupling only 
  $G^{00}$ and $\bfG^{10}$ will be finite.}

\begin{table}[t]
\caption{Spin dependence and parity properties of the four components in the expansion of the single electron Green function $\bfG$.}
\label{tab-GFs}
\begin{tabular}{c||c|c|c}
\hline
Green function & spin reversal & space reversal & time reversal\\\hline
$G^{00}$ & even & even & even\\
$G^{01}$ & even & odd & odd \\\hline
$\bfG^{10}$ & odd & even & odd \\
$\bfG^{11}$ & odd & odd & even\\\hline
\end{tabular}
\end{table}

{This static interaction can for clarity and consistency with earlier literature \cite{Heisenberg26,Liechtenstein1987} on bi-linear exchange couplings also be expressed in a discrete atomic site or lattice formalism. The 
  $\bfrho$ and $\bfrho'$ integrations in Eq.~(\ref{eq-Tpq})  are then taken to be over atomic sites $i$ and $j$ and the local interactions $\Xi_p(\bfr,\bfrho)$ are assumed to be on-site only. In order to perform these integrals we expand all quantities in local orbitals, e.g.~spherical or tesseral harmonics, which render all quantities to be matrices in this orbital space, although the local interaction $\Xi^p_i$ is usually taken to be diagonal. Then we get a lattice representation of Eqs.~(\ref{eq-Tpq}) or (\ref{eq-Kpq}) as
\begin{align}
	T^{pq}_{ij}=&\frac{1}{2^{\delta_{ps}+\delta_{qs}}} {\rm sp} \, \tr\,\im \int f(\dote{})
	\Xi^p_i\bfsigma_p\,\bfG_{ij}
			\Xi^p_j\bfsigma_q
			\bfG_{ji}
	d\dote{}\,,
\end{align}
where the trace is now over both spin (${\rm sp}$) and orbital ($\tr$) space. 
In this matrix formalism the Green function is a matrix over both spin and orbitals. Then when we decompose it in the way of Eq.~(\ref{eq-2index}) each term is still a matrix over orbitals.
This fact lead to that the decomposed Green functions are not anymore simply even or odd under change of direction or equivalently site exchange. Instead, in case of a real basis we  have that
\begin{align}
	G^{00}_{ij}=&\{G^{00}_{ji}\}^T\nonumber\\
	G^{01}_{ij}=&-\{G^{01}_{ji}\}^T\nonumber\\
	\bfG^{10}_{ij}=&\{\bfG^{10}_{ji}\}^T\nonumber\\
	\bfG^{11}_{ij}=&-\{\bfG^{11}_{ji}\}^T\,,\label{eq-G2index-symm}
\end{align}
where the matrix transpose is over the orbitals. For general complex orbitals this relation will depend on the choice of basis, therefore we restrict to real basis in this paper and the expressions below for the interaction parameters are only valid for this special case.}

\subsection{Lattice-lattice coupling}
\label{aapp-pp}
{
\begin{widetext}
Applying the introduced decomposition of the Green function to the interatomic force constant presented in Eq.~(\ref{eq:dynspinlatt}) 
we obtain the form
\begin{align}
\Phi_{ij}=&
	-\frac{4}{\pi} \tr
	\im
	\int f(\dote{})
		\Bigl(
				\bflambda_i G^{00}_{ij}	\bflambda_j G^{00}_{ji}
			+
				\bflambda_i G^{01}_{ij}	\bflambda_j G^{01}_{ji}
			+
				\bflambda_i \bfG^{10}_{ij}	\bflambda_j \bfG^{10}_{ji}
			+
				\bflambda_i \bfG^{11}_{ij}	\bflambda_j \bfG^{11}_{ji}
		\Bigr)
	d\dote{}\,,
\end{align}
\end{widetext}
where the products between the Green functions $\bfG^{10}$ and $\bfG^{11}$ in the third and fourth term, respectively, should be considered as scalar products.
The presence of the spin-dependent components shows that also the spin texture in the material can have a crucial influence on the lattice-lattice coupling in the material, which lead to the well-known fact that the atomic forces will be spin dependent for a magnetic system.
}

\subsection{Spin-spin coupling}
\label{aapp-ss}
{As displayed in Eq.~(\ref{eq-dyn-ss})
the indirect spin-spin exchange can be partitioned into three contributions: isotropic Heisenberg, anisotropic Dzyaloshinskii-Moriya  and Ising interactions.
By application of the Green function decomposition introduced, we find that these three interactions in the static limit can be written as
\begin{widetext}
\begin{subequations}
\label{eq-genDss}
\begin{align}
J_{ij}=&
	-\frac{1}{2\pi} \tr
	\im
	\int f(\dote{})
		\Bigl(\nu_i G^{00}_{ij}\nu_j G^{00}_{ji}
			+
			\nu_i G^{01}_{ij}\nu_j G^{01}_{ji}
			-
			\nu_i \bfG^{10}_{ij}\cdot\nu_j \bfG^{10}_{ji}
			-
			\nu_i \bfG^{11}_{ij}\cdot\nu_j \bfG^{11}_{ji}		
	    \Bigr)
	    d\dote{}
	,
\label{eq-genJ}
\\
\bfD_{ij}=&
	-\frac{2}{\pi} \tr
	\re
	\int f(\dote{})
		\Bigl(
			\nu_i G^{00}_{ij}\nu_j \bfG^{11}_{ji}
			+
			\nu_i G^{01}_{ij}\nu_j \bfG^{10}_{ji}
		\Bigr)
	d\dote{}
	,
\label{eq-genDM}
\\
\mathbb{I}_{ij}=&
	-\frac{2}{\pi} \tr
	\im
	\int f(\dote{})	
		\Bigl(
			\nu_i \bfG^{10}_{ij}	\nu_j \bfG^{10}_{ji}
			+
			\nu_i \bfG^{11}_{ij}	\nu_j \bfG^{11}_{ji}
		\Bigr)
	d\dote{}
	.
\label{eq-genI}
\end{align}
\end{subequations}
\end{widetext}}
First, it is important to notice that the three contributions are given as a scalar ($J$), vector ($\bfD$), and a dyad ($\mathbb{I}$), as would be an expected partitioning of a second rank tensor. {These interactions are closely related to other expressions for $J$ and $\bfD$ in the literature \cite{Liechtenstein1987,Udvardi2003,Ebert2009}, now expressed in decomposed Green functions. 
Second, we notice that since $G^{00}$ or $\bfG^{10}$ are always present in a magnetic systems, the Dzyaloshinskii-Moriya interaction can be finite only when either $\bfG^{11}$ or $G^{01}$ do not vanish. As mentioned above these two functions vanish in the absence of spin-orbit coupling for topologically trivial materials in equilibrium. 
Third, it is important to observe that the Ising interaction in its most general form, as here, is represented by a dyad and due to its first term can be non-vanishing also in 
the non-relativistic limit without spin-orbit coupling.}

\subsection{Spin-lattice coupling}
\label{aapp-sp}
Finally, the spin-lattice interactions in the static limit $T^{cs}_{ij}$ derived from Eq.~(\ref{eq-genDcs}), can in terms of the four Green functions in Table \ref{tab-GFs} be written as
\begin{widetext}
\begin{subequations}
\begin{align}
T^{cs}_{ij}=&
	-\frac{4}{\pi}
	 \tr \im
	\int
		f(\dote{})	\Bigl(
		\bflambda_i
			G^{00}_{ij}	\nu_j\bfG^{10}_{ji}
			+
			\bflambda_iG^{01}_{ij}	\nu_j\bfG^{11}_{ji}
			-
			i\bflambda_i\bfG^{10}_{ij}\times\nu_j\bfG^{11}_{ji}
		\Bigr)	
	d\dote{}
	.
\end{align}
\label{eq-TscTcs}
\end{subequations}
\end{widetext}
It is easily seen that the tensor $T^{sc}$ is related to this tensor by the transpose
\begin{align}
	\left\{T^{sc}_{ij}\right\}^{\alpha\beta}
	= \left\{T^{cs}_{ji}\right\}^{\beta\alpha}\,,\label{KijT}
\end{align}
with the explicit Cartesian tensor components $\alpha$ and $\beta$.
The $T^{cs}$ tensor interactions can further be partitioned into two independent terms,
$T^{cs}=\mathcal{S}+\mathcal{A}$, with
\begin{subequations}
\begin{align}
\mathcal{S}_{ij}=&
	-\frac{4}{\pi}
	 \tr \im
	\int
		f(\dote{})
		\Bigl(
				\bflambda_i G^{00}_{ij}\nu_j\bfG^{10}_{ji}
			+
				\bflambda_i G^{01}_{ij}\nu_j\bfG^{11}_{ji}
		\Bigr)
	d\dote{}
	,
\\
\mathcal{A}_{ij}=&
	-\frac{4}{\pi}
	 \tr \re
	\int
		f(\dote{})
		\bflambda_i
		\bfG^{10}_{ij}\times\nu_j\bfG^{11}_{ji}
	d\dote{}
	.
\end{align}
\label{eq-AS}
\end{subequations}

Then it is noteworthy that the $\mathcal{S}$ interaction is even in the spin-orbit coupling strength while the $\mathcal{A}$ in contrast is odd. Hence, for systems with weak spin-orbit coupling, the first interaction is expected to dominate if it is allowed by symmetry.
It is straight-forward from Eq.~(\ref{eq-AS}) to verify that the first term is symmetric with respect to site exchange $\mathcal{S}_{ij}=\mathcal{S}_{ji}$ while the second is anti-symmetric $\mathcal{A}_{ij}=-\mathcal{A}_{ji}$, by using the relations for the decomposed Green functions of Eq.~(\ref{eq-G2index-symm}).

\section{Symmetries}
\label{sec-symmetries}

We want to study the symmetry of the spin-lattice part of the static interaction Eq.~\eqref{eq-effHdiscrete}, i.e.~the heterogenous part
 \begin{align}
   \mathcal{H}_\text{MQ}^\mathrm{sl}=&-\frac{1}{2}\sum_{ij}\,\left\{ {\bf Q}_i\,\cdot T^{cs}_{ij}\cdot {\bf M}_j
   +{\bf M}_i\,\cdot T^{sc}_{ij}\cdot {\bf Q}_j\right\}\,.\label{Ene}
 \end{align}
The fact that the two quantities entering this bi-linear form have different symmetries might cause some confusion. The lattice distortion ${\bf Q}_i$ is even under time reversal $\theta$ but odd under space inversion $\iota$ while the magnetic moment ${\bf M}_i$ is invariant with respect to space inversion but change sign under operation of time reversal. Hence since the interaction energy is scalar, the interaction coefficients
have to be odd under both space inversion and time reversal
which single out the heterogenous bi-linear spin-lattice interaction compared to the homogenous bi-linear spin-spin $\{T^{ss}\}$ and lattice-lattice $\{T^{cc}\}$ interactions, that are both invariant under these operations. However, when accepting this difference there is nothing that forbid such heterogenous interactions, as will be demonstrated below,  first through derivation of explicit expressions for these interaction parameters and then by considering the symmetry of the interactions. The odd time reversal property is simply stated as 
$\theta T^{cs}_{ij} = -T^{cs}_{ij}$ while the space inversion has to be discussed in more details below.
   
First we notice for each pair $\{ij\}$ of sites we have four interaction terms 
\begin{align}
{\bf Q}_i \cdot T^{cs}_{ij} \cdot{\bf M}_j + {\bf Q}_j\cdot T^{cs}_{ji}\cdot {\bf M}_i+
{\bf M}_i\cdot T^{sc}_{ij} \cdot{\bf Q}_j+ {\bf M}_j\cdot T^{sc}_{ji}\cdot {\bf Q}_i\,.	
\end{align}
Then from the relation (\ref{KijT}) 
\begin{align}
	{\bf Q}_i \cdot T^{cs}_{ij}\cdot{\bf M}_j+{\bf M}_j\cdot T^{sc}_{ji}\cdot{\bf Q}_i
	= 2{\bf Q}_i\cdot T^{cs}_{ij}\cdot {\bf M}_j 
	\,,
\end{align}
so the total spin-lattice interaction written as a sum over pairs becomes
\begin{align}
	\mathcal{H}_\mathrm{MQ}^\mathrm{sl}=&-\sum_{\{ij\}}\sum_{\alpha\beta}\left(\{T^{cs}_{ij}\}^{\alpha\beta}Q^\alpha_i{M}^\beta_j+\{T^{cs}_{ji}\}^{\alpha\beta}Q^\alpha_j{M}^\beta_i \right)\,.
	\label{Eint2}
\end{align}
{\begin{widetext}
Now we can decompose the pair interaction into a part that is symmetric $\mathcal{S}_{ij}$ and one that is antisymmetric $\mathcal{A}_{ij}$ with respect to interchange of sites, i.e.\ 
with ${T}^{cs}_{ij}=\mathcal{S}_{ij}+\mathcal{A}_{ij}$ we have that ${T}^{cs}_{ji}=\mathcal{S}_{ij}-\mathcal{A}_{ij}$. Then Eq.~(\ref{Eint2}) becomes
\begin{align}
	\mathcal{H}_\mathrm{MQ}^\mathrm{sl}=&-\sum_{\{ij\}}\sum_{\alpha\beta}\left[\{\mathcal{S}_{ij}\}^{\alpha\beta}\left(Q^\alpha_i{M}^\beta_j+Q^\alpha_j{M}^\beta_i\right)+\{\mathcal{A}_{ij}\}^{\alpha\beta}\left(Q^\alpha_i{M}^\beta_j-Q^\alpha_j{M}^\beta_i\right) \right]\nonumber\\
	=&-\sum_{\{ij\}}\big\{{\bf Q}_i\cdot\mathcal{S}_{ij}\cdot{\bf M}_j+{\bf Q}_j\cdot\mathcal{S}_{ij}\cdot{\bf M}_i+{\bf Q}_i\cdot\mathcal{A}_{ij}\cdot{\bf M}_j -{\bf Q}_j\cdot\mathcal{A}_{ij}\cdot{\bf M}_i\big\} \label{pairinter}
\end{align}
In contrast to the homogeneous bi-linear interactions these interaction parameters $\mathcal{S}_{ij}$ and $\mathcal{A}_{ij}$ are both general rank two tensors in 3D space and can hence both be decomposed into three  contributions, scalar ($S_{ij}$ and $A_{ij}$), vector (${\bf S}_{ij}$ and ${\bf A}_{ij}$) and symmetric second rank tensor interactions (${S}^{(2)}_{ij}$ and ${A}^{(2)}_{ij}$).
By first decomposing these interaction tensors into  symmetric and anti-symmetric parts $\mathcal{S}_{ij}=\,_s\mathcal{S}_{ij}+\,_a\mathcal{S}_{ij}$ with respect to exchange of components, the interaction energy can be expressed as
\begin{align}
		\mathcal{H}_\mathrm{MQ}^\mathrm{sl}=-\frac{1}{2}\sum_{\{ij\}\alpha\beta} &\left\{_s\mathcal{S}_{ij}^{\alpha\beta}\left( Q^\alpha_i{M}^\beta_j+Q^\beta_i{M}^\alpha_j+Q^\alpha_j{M}^\beta_i+Q^\beta_j{M}^\alpha_i\right)\right.+\,
		_s\mathcal{A}_{ij}^{\alpha\beta}\left( Q^\alpha_i{M}^\beta_j+Q^\beta_i{M}^\alpha_j-Q^\alpha_j{M}^\beta_i-Q^\beta_j{M}^\alpha_i\right)+\nonumber\\
		+&\,_a\mathcal{S}_{ij}^{\alpha\beta}\left(Q^\alpha_i{M}^\beta_j-Q^\beta_i{M}^\alpha_j+Q^\alpha_j{M}^\beta_i-Q^\beta_j{M}^\alpha_i \right)+	\,
		_a\mathcal{A}_{ij}^{\alpha\beta}\left.\left(Q^\alpha_i{M}^\beta_j-Q^\beta_i{M}^\alpha_j-Q^\alpha_j{M}^\beta_i+Q^\beta_j{M}^\alpha_i \right)\right\}=\nonumber\\
	 =-\sum_{\{ij\}} &\left\{S_{ij}\left( {\bf Q}_i\cdot{\bf M}_j+{\bf Q}_j\cdot{\bf M}_i\right)\right.
		+A_{ij}\left( {\bf Q}_i\cdot{\bf M}_j-{\bf Q}_j\cdot{\bf M}_i\right)
		+{\bf S}_{ij}\cdot\left({\bf Q}_i\times{\bf M}_j+	{\bf Q}_j\times {\bf M}_i\right)
	+{\bf A}_{ij}\left.\cdot\left({\bf Q}_i\times{\bf M}_j-{\bf Q}_j\times {\bf M}_i  \right)+\ldots\right\}\nonumber\\
	=-\sum_{ij}&\left({\bf Q}_i\cdot(\mathcal{S}_{ij}+\mathcal{A}_{ij}\right)\cdot{\bf M}_j
	=-\sum_{ij}\left({S}_{ij}+{A}_{ij}\right){\bf Q}_i\cdot{\bf M}_j
	-\sum_{ij}\left({\bf S}_{ij}+{\bf A}_{ij}\right)\cdot {\bf Q}_i\times{\bf M}_j
	-\sum_{ij}{\bf Q}_i\cdot\left({S}_{ij}^{(2)}+{A}_{ij}^{(2)}\right)\cdot{\bf M}_j\,,\label{sitesum}
\end{align}
\end{widetext}
where the dots refer to the for moment neglected second rank contributions and note that in the last line we do the full site sum again.  Scalar and vector interactions have been introduced in line with conventions.  The scalar interactions are related to the trace of the symmetric part of the tensors, while the vector interactions are the dual form of the anti-symmetric part of the tensors. So for the symmetric tensor $\mathcal{S}_{ij}$ we decompose it in terms of
\begin{align}
S_{ij}&=\frac{1}{3}\mathrm{Tr}\,_s\mathcal{S}_{ij}\,,
\end{align}
and
\begin{align}
	{S}_{ij}^\gamma &={\bf S}_{ij}\cdot \hat{\gamma}=\frac{1}{2}\sum_{\alpha\beta}\,\epsilon_{\alpha\beta\gamma}\,_a\mathcal{S}_{ij}^{\alpha\beta}\,.
\end{align}
where $\epsilon_{\alpha\beta\gamma}$ is the anti-symmetric Levi-Civita symbol and $\hat{\gamma}$ is the unit vector along Cartesian axis $\gamma$.
Finally the second rank tensor interactions $S_{ij}^{(2)}$ 
is given
as
\begin{align}
	S_{ij}^{(2)}&=\,_s\mathcal{S}_{ij}-S_{ij}{\bf 1}\,,
\end{align} 
where ${\bf 1}$ is the 3D unit matrix.}

In order to discuss the symmetry under space inversion, let us consider that the inversion operation $\iota$  brings site $i$ to an equivalent site $i'$ and correspondingly for site $j$.
{In Appendix \ref{App-Symm} it is shown that in this case both spin-lattice interaction tensors are indeed odd under space inversion, i.e.,
\begin{align}
	\iota\mathcal{S}_{ij}=&-\mathcal{S}_{i'j'}\nonumber\\
	\iota\mathcal{A}_{ij}=&-\mathcal{A}_{i'j'}\,.
\end{align}
For the special case where there exists a center of inversion at the bond center in between sites $i$ and $j$, inversion brings site $i$ to site $j$ and
\begin{align}
	\iota\mathcal{S}_{ij}=&-\mathcal{S}_{ji}=-\mathcal{S}_{ij}\nonumber\\
	\iota\mathcal{A}_{ij}=&-\mathcal{A}_{ji}=\mathcal{A}_{ij}\,,
\end{align}
Hence in this case the interaction tensor $\mathcal{S}_{ij}$ has to vanish.
If instead there exist a bond center invariant under the combined operation of space inversion and time reversal $\iota\theta$, then instead
\begin{align}
	\iota\theta\mathcal{S}_{ij}=&\mathcal{S}_{ji}=\mathcal{S}_{ij}\nonumber\\
	\iota\theta\mathcal{A}_{ij}=&\mathcal{A}_{ji}=-\mathcal{A}_{ij}\,,
\end{align}
i.e.~$\mathcal{A}_{ij}$ has to vanish.}

{
This reminds about the fact that 
the Dzyaloshinskii-Moriya interaction  ${\bf D}_{ij}$ of Eq.~(\ref{eq-genDM}), also vanishes if there is an inversion symmetry at the bond center.
However, a difference is that the Dzyaloshinskii-Moriya interaction is even under the inversion per se. It is the asymmetry under site exchange which makes it vanish,
$\iota{\bf D}_{ij}={\bf D}_{ji}=-{\bf D}_{ij}$.}

{In the full magnetic symmetry group the elements generally consist of combined operations, e.g.~rotations and inversion or rotations and time reversal etc as illustrated in the examples below. The rotational part of this operation behaves as expected, either on the full interaction tensor or the scalar and vector interactions in its decomposition, while as noted both inversion and time reversal operations are odd for the spin-lattice interaction. }

Finally it is important to remember that for the heterogenous spin-lattice interaction the inter-site exchange symmetry is unrelated to the symmetry of the tensor. So the interaction contribution that is symmetric in site exchange, $\calS_{ij}$, contributes both to the scalar interaction $\bfQ_i\cdot\bfM_j$ as well as the cross product interaction
$\bfQ_i\times\bfM_j$. This is in contrast to the homogeneous spin-spin interaction where the interaction symmetric in sites, e.g.~Heisenberg, only contributes to the symmetric scalar interaction $\bfM_i\cdot\bfM_j$ etc.
Anyhow we have chosen to differ between the two contributions as they behave differently with the strength of the spin-orbit coupling. The symmetric interaction $\calS_{ij}$ exists also in absence of spin-orbit coupling while the anti-symmetric $\calA_{ij}$ is linear in a weak spin-orbit coupling strength as shown by Eq.~(\ref{eq-AS}).

\section{Examples}
\label{sec-examples}

\subsection{Numerical Details}
The bilinear couplings \eqref{eq-Dpq} are implemented in our real space tight binding code \footnote{\textit{CAHMD} - classical atomistic Heisenberg magnetization dynamics, 2013. A computer program package for atomistic magnetization dynamics simulations. Available from the authors; electronic address: danny.thonig@physics.uu.se.}. Here, we solve the non-orthogonal eigenvalue problem ${\cal{H}}\psi = \varepsilon{\cal{O}}\psi$ where $\psi$ is a linear combination of atomic orbitals (LCAO ansatz) within a $sp_3d_5$ orbital basis set. The Hamiltonian $\mathcal{H}_0$ and the overlap matrix $\cal{O}$ are build up from the Slater-Koster scheme \cite{Slater1954}, where the Slater-Koster parameter are consider distance dependent according to the formalism of Mehl et al. \cite{Mehl1996,Cohen1994}. The full Hamiltonian ${\cal{H}}={\cal{H}}_0+{\cal{H}}_{\mathrm{soc}}+{\cal{H}}_{\mathrm{mag}}$ includes also spin-orbit coupling ${\cal{H}}_{\mathrm{soc}}=\xi \vec{L}\cdot\vec{S}$ and magnetic exchange splitting ${\cal{H}}_{\mathrm{mag}}=\frac{I}{2} \vec{M} \cdot \vec{S}$, respectively. Both the spin-orbit coupling parameter $\xi$ and the Stoner excitation energy $I$ are obtained from fitting of the electronic structure to ab-initio band structures obtained from a full-relativistic multiple scattering Green's function method (Korringa-Kohn-Rostoker method, KKR)\cite{Zabloudil05}.  $\vec{M}=m \vec{e}_s$ is the spin magnetic moment. Magnetic moment rotations come from a unitary transformation of the Hamiltonian with relativistic rotation matrices $\cal{R}$, consisting of rotations in spin and orbital space \cite{Lindner84}. Variations of the magnetic moment $\vec{e}_s=\vec{e}_s(\theta,\phi)$ are addressed by $\nicefrac{\partial \cal{H}}{\partial \theta_i}$ and $\nicefrac{\partial \cal{H}}{\partial \phi_{i}}$. A local approximation for $\bflambda_i$ is used by the derivative of the Hamiltonian $\nicefrac{\partial \cal{H}}{\partial \vec{Q}_i}$ due to lattice degrees of freedom $\vec{Q}_i$, obtained from Ref.~\cite{Dziedzc06}. In the simulations we focus on low dimensional clusters of Fe, e.g., chains, with periodic boundary conditions, where the tight binding parameters are from Refs.~\cite{Mehl2008pp,Schena10,Thonig2017}.

Since pure spin and lattice exchange couplings \cite{Pajda2001,Bottcher2012,Diaz-Sanchez2007,Luo2010} are already well understood, we will focus in the following only on the bilinear spin-lattice coupling mechanism, and then especially the influence on the lattice from the spin order.

\subsection{Double anti-ferromagnetic lattice}
\label{ssec-dafm}
It is discussed in literature \cite{Sjostedt2002} that the magnetic ground state in fcc Fe is double anti-ferromagnetic. It is collinear with all moments along, say, the $\hat{z}$-direction, where the variations along, say, the $\hat{x}$-direction, is $\uparrow\uparrow\downarrow\downarrow$ and translations of this unit cell (cf. Fig.~\ref{fig:DAFM}). The symmetry group for this spin structure is $\{e,\iota,\theta t_2,\iota\theta t_2\}\otimes T$, where $T=\{nt_4; n\in\mathbb Z\}$ is all pure translations of the unit cell and $t_2$ is a non-trivial translation by two sites. $e$, $\iota$, and $\theta$ are the identity, inversion and the spin (time) reversal operator, respectively. Note that this choice of symmetry group is quantization axis free and, consequently, suitable for non-relativistic treatment. The inversion center can be chosen as in between atoms $1$ and $2$ or equivalently in between atoms $3$ and $4$ (see Section \ref{sec-symmetries}).

Without spin-orbit coupling, rotational variation of the magnetic moment $\delta\theta,\delta\phi$ makes $\nu(\bfr,\bfrho)$ in Eq.~\eqref{eq-genDss} proportional to the Pauli matrices $\sigma_x, \sigma_y$. Hence, they do not contribute to spin-lattice coupling due to the spin-diagonal from of the Green's function. It turns out that for the double anti-ferromagnetic structure $T^{cs}_{ij}$ is related to longitudinal fluctuations of the magnetic moments, which is proportional to $\sigma_z$ (Fig.~\ref{fig:DAFM}). To apply the group symmetry analysis, it is useful to treat the couplings to be at the center of the bonds between atoms (cf. Fig.~\ref{fig:DAFM}). Here, the symmetric scalar interactions $S_{ij}$ vanish at the bond centers 1-2 and 3-4, due to inversion. However in between 4-1 and 2-3 they can exist and are related by $\theta t_2$, i.e.~$S_{41}=-S_{23}=s$. So there will be forces $\vec{F}_i=-\nicefrac{\partial \cal{H}^{\text{sl}}_{MQ}}{\partial \vec{Q}_i}$ on all four atoms
 \begin{align}
 	F_1&=-S_{14}m_4=+s\nonumber\\
 	F_2&=-S_{13}m_3=-s\nonumber\\
 	F_3&=-S_{32}m_2=+s\nonumber\\
 	F_4&=-S_{41}m_1=-s\,,
 \end{align}
which leads to a dimerization; atoms 1 and 2, respectively, 3 and 4, move towards each other (cf. Fig.~\ref{fig:DAFM} (a) - black arrow). This was also approved numerically (Fig.~\ref{fig:DAFM} (a)) by comparing different collinear magnetic textures, a ferromagnetic (FM), anti-ferromagnetic (AFM), and double anti-ferromagnetic (DAFM) structure.

\begin{figure}
\centering
 \includegraphics[width = 0.9\columnwidth]{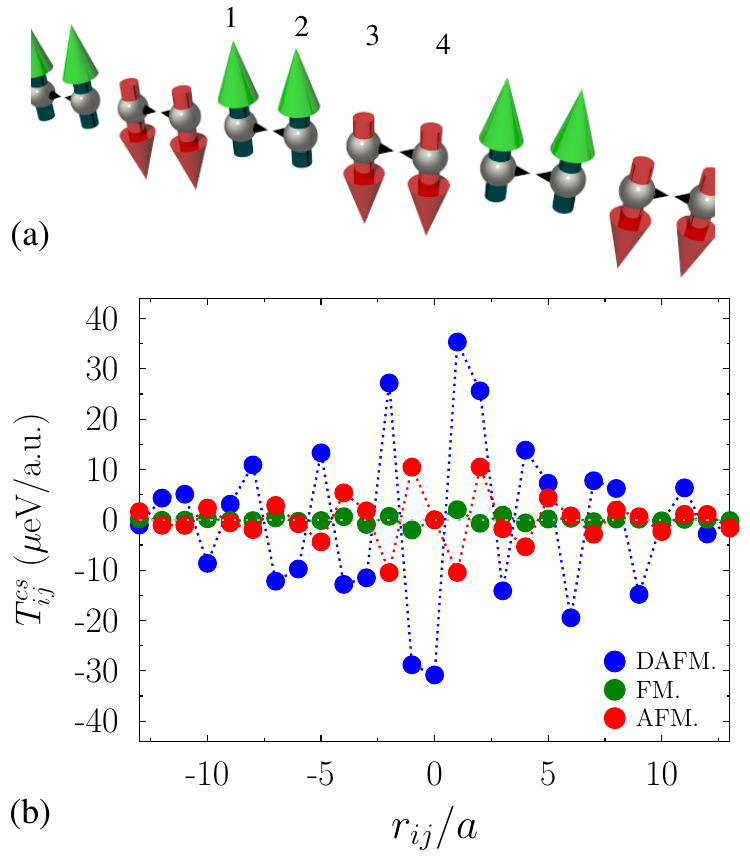}
\caption{(Color online) \textit{(a)} Magnetic moment structure (bold arrows) and related forces (black arrows) coming from bilinear spin-lattice coupling for the double antiferromagnetic. Atoms are indicated by gray balls. The color of the magnetic moments indicate the orientation ($\hat{z}$ - green arrow; $-\hat{z}$ - red arrow). \textit{(b)} Bilinear spin-lattice coupling vs. distance along the variation $\uparrow\uparrow\downarrow\downarrow$ for different magnetic states: ferromagnetic (FM, green dots), anti-ferromagnetic (AFM, red dotes), and double anti-ferromagnetic (DAFM, blue dotes). }
\label{fig:DAFM}
\end{figure}

Note that the magnetic moment length is set to $\unit[0.001]{\mu_B}$ for a proper ground state description. $T^{cs}_{ij}$ scales linear the moment length; thus the nearest neighbour coupling  $T_{NN}$ is $\approx\unit[10]{eV}$ for the magnetic moment length $m=\unit[2.3]{\mu_B}$ for Fe. In the first two cases, say FM and AFM, the exchange $T^{cs}_{ij}$ is antisymmetric around zero and, consequently, no net-force exists. However, it effects the dynamics of the spin and lattice degree of freedom. Oscillations occur for the AFM structure which is linked to the alternating spin state. $T^{cs}_{ij}$ in the DAFM is not antisymmetric around the origin, but around the bond center, described in our symmetry analysis. This originates an alternating finite force of $F=\unit[0.39]{\mu eV/a.u.}$ ($F=\unit[1.33]{eV/a.u.}$ for finite moment) between the atoms and causes dimerization of the atoms.

\subsection{Planar spin density waves}
\label{ssec-planar}

\begin{figure}
\centering
 \includegraphics[width = 0.9\columnwidth]{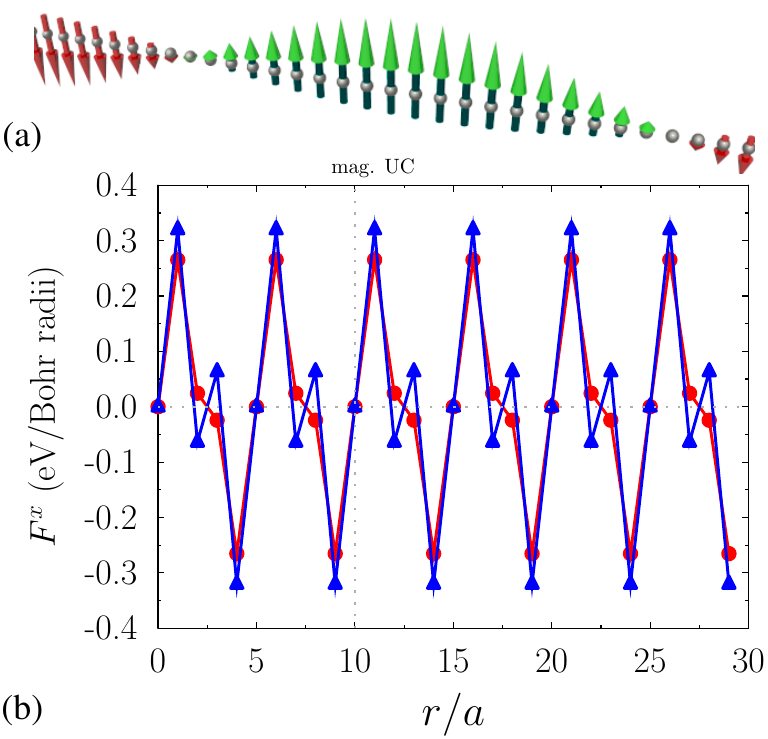}
\caption{\label{sSDW}(Color online) a) Magnetic moment structure (bold arrows) for the planar spin density wave. b) Lattice forces at different position in a sinusoidal spin density wave calculated from the force related to the bilinear spin-lattice coupling (blue triangles) and from the Hellmann-Feynman theorem (red dots). The oscillation period extends over 10 atoms, indicated by the vertical dotted line. Lines are edit to guide the eye. }
\label{fig:sSDW}
\end{figure}

In the previous case we kept the crystal and spin structure to be simple. If we extend the two magnetic structures to an infinite spiral, represented by 
\begin{align}
 \vec{M}_i=M_z\hat{z}\cos(qx_i)+M_e\hat{e}\sin(qx_i)\,
 \label{eq:spiral}
\end{align}
where $x_i$ is the $x$-component of the position of atom $i$, $\vec{r}_i$, $q$ is the magnitude of the wave vector $\vec{q}=q\hat{x}$ and $\hat{e}$ is either \text{i)} $\hat{x}$ or \textit{ii)} $\hat{y}$. For $M_e=0$ the magnetic structure would correspond to a sinusoidal spin density waves (sSDW) (Fig.~\ref{sSDW} a). Here, two phases are possible, either with a belly or node at $x_0=0$, respectively. We notice that the symmetry groups for the sinusoidal spin density wave is for the belly $\left\{e,c_{2z},\theta c_{2x}, \theta c_{2y}\right\}\times\left\{e,\iota\right\}$ and for the node $\left\{e,c_{2z},\theta c_{2x}, \theta c_{2y}\right\}\times\left\{e,i\theta\right\}$. Here, $c_{n\nu}$ defines the $n$-fold rotation axis along $\nu$. Note that the sinusoidal magnetic structure is invariant with respect to $\iota$ or $\iota\theta$ for the belly or nodal type, respectively. 

Let us focus on the symmetric scalar interaction for the belly sinusoidal SDW with a node at $qx=0$ and a maximum at $qx=\pi/2$. Thus (not shown here), the symmetric scalar interaction behaves as $s_j=s\sin q(x_j+d/2)$ and the force at atom $j$ due to its nearest neighbour interactions are, 
\begin{align}
	F_j&=-\left(S_{jj-1}M_{j-1}+S_{jj+1}M_{j+1}\right)\nonumber\\
	&=-2sM\sin 2qx_j \cos qd/2\,,
	\label{sinus-force}
\end{align}
where $d$ is the distance between two atoms. These forces oscillate with $2q$, however they disappear for $qd/2=\pi/2$, i.e.~for $qd=\pi$ which correspond to a commensurate AFM, where the variation of magnetic moments disappear, i.e.~$\vec{m}_j=0$. Note that we recounter the double layered AF for $qd=\pi/2$ if the phase shift the sSDW with $\phi=-3d/2$. This periodicity is also recovered by our numerical method (\ref{fig:sSDW} b).

Note that the calculations are done for a finite magnetic moment of $\unit[2.23]{\mu_B}$ and all neighbours contribute to the summation needed to get the force from the bilinear coupling term. This results in slight variation from the $\sin$-like behavior of the force observed from group symmetric analysis. The periodicity of $2q$, however, was reproduced. The obtained forces are in good agreement with the forces obtained directly from Hellmann-Feynman theorem. Disparities are due to higher orders exchange couplings that are included in the Hellmann-Feynman force as well as due to long-range exchange. 

\subsection{Cycloidal and helical spin density wave}
\label{ssec-helical}

Continuing the discussion about spiral spin configurations \eqref{eq:spiral}, we set $M_z=M_e=M$, which correspond to either a cycloidal spiral $\hat{e}=\hat{x}$ or helical spiral $\hat{e}=\hat{y}$, respectively, with the symmetry groups $\left\{e,c_{2z},i\theta c_{2x}, i\theta c_{2y}\right\}$ for the cycloid  and $\left\{e,c_{2z},\theta c_{2x}, \theta c_{2y}\right\}$ for the helix state.

\paragraph{Cycloid} For a general position of an atom $j$ at $\vec{r}_j$, the bond center to the nearest neighbour is conserved by the group $\left\{e,\iota\theta c_{2y}\right\}$ and allows both a scalar $T^{cs;s}_{jj\pm 1}$ and a vectorial coupling along $y$, $\vec{T^{cs;v}}_{jj\pm1}\times\hat{y}=0$. We assume the spiral to be commensurate and point to the atomic position $\vec{r}_n$ such that $qx_n = \nicefrac{\pi}{2}$. Caused by the symmetry operation $\theta c_{2z}$, the sign of the nearest neighbour scalar interaction, $T^{cs;s}_{nn+1}=-T^{cs;s}_{nn-1}$, changes, while the nearest neighbour vector interaction does not $\vec{T}^{cs;v}_{nn+1}=\vec{T}^{cs;v}_{nn-1}$; it behaves opposite as around the point $\vec{r}_0$. This can be explained only by an oscillatory behaviour of the interaction parameters along the wave vector. 

There are two possibilities, either the interactions are symmetric (see section \ref{sec-symmetries}), $S_{jj+1}$ and $\vec{S}_{jj+1}$, or anti-symmetric, $A_{jj+1}$ and $\vec{A}_{jj+1}$. This gives rise to the following forces on the atom at $\vec{r}_j$ arising from nearest neighbour interactions due to last part in Eq.~\eqref{sitesum}:
\begin{subequations}
\label{eq-forcecyc}	
\begin{align}
	\vec{F}_j^{Ss}&=-\left(S_{jj-1}\vec{M}_{j-1}+S_{jj+1}\vec{M}_{j+1}\right)\nonumber\\
	&=-sM\vec{e}(qx_j)\left[2\cos(qd)-1\right]\cos(qd/2)
	\label{eq-forcecycss}	
\\
	\vec{F}_j^{Sv}&=-\left(\vec{M}_{j-1}\times\vec{S}_{jj-1}+\vec{M}_{j+1}\times\vec{S}_{jj+1}\right)\nonumber\\
	&=s_yM\left\{\left[2\cos(qd)-1\right]\vec{e}(qx_j)+\hat{z}\cos(qd)\right\}\cos(qd/2)
	\label{eq-forcecycsv}
\\
	\vec{F}_j^{As}&=-\left(A_{jj-1}\vec{M}_{j-1}+A_{jj+1}\vec{M}_{j+1}\right)\nonumber \\
	&=-aM\vec{e}(qx_j)\left[2\cos(qd)-1\right]\cos(qd/2)
	\label{eq-forcecycas}
\\
	\vec{F}_j^{Av}&=-\left(\vec{M}_{j-1}\times\vec{A}_{jj-1}+\vec{M}_{j+1}\times\vec{A}_{jj+1}\right)\nonumber\\
	&=-a_yM\left\{\vec{e}(qx_j)(2\cos(qd)+1)+\hat{z}\right\}\sin(qd/2),
	\label{eq-forcecycav}
\end{align}
\end{subequations}
where $\vec{e}(qx_j)=\left\{\hat{z}\cos(2qx_j)+\hat{x}\sin(2qx_j)\right\}$. Here, $s$ and $a$ are the magnitude of the oscillating  antisymmetric and symmetric, scalar and vectorial couplings. The symmetric scalar force \eqref{eq-forcecycss} does not vanish in the limit $q\rightarrow0$. The symmetric vector \eqref{eq-forcecycsv} and anti-symmetric vector force \eqref{eq-forcecycav} vanishes in the limit $q\rightarrow0$, but has otherwise in addition to the oscillations also a constant term in $\hat{z}$-direction, where the $z$-component goes as $\{(2\cos qd+1)\cos(2qx_j)+1\}\sin qd/2$.

\paragraph{Helix} For the helical spin spiral state, the anti-symmetric interaction has two non-vanishing components, since $\theta c_{2x} \vec{A}_{ij}=\vec{A}_{ij}$, and from the symmetry relations at $qx=0$ and $qx=\pi/2$ they have to exhibit also an oscillatory behaviour. This leads to a force at atom $j$ as
\begin{align}
		\vec{F}_j^{Av}
		&=-\left(\vec{M}_{j-1}\times\vec{A}_{jj-1}+\vec{M}_{j+1}\times\vec{A}_{jj+1}\right)\nonumber \\
		&=-M\left\{(\cos qd +1)(a_y+a_z)-a_y\right\}\sin(qd/2)\sin(2qx_j)\hat{x}\,,
\end{align}
which is also purely oscillatory.

\begin{figure}
\centering
 \includegraphics[width = 0.9\columnwidth]{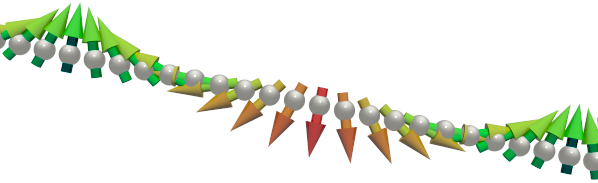}
\caption{(Color online) Magnetic moment structure (bold arrows) for the cycloidal spin wave. }
\label{fig:Schematic}
\end{figure}

To summarize, for 
the magnetic textures with wave vector $\vec{q}$ discussed in Secs.~\ref{ssec-planar} and \ref{ssec-helical}, 
we observe an oscillating force with the double wave vector. In particular for the cycloidal spin wave, we obtained a constant force in addition to the oscillating force. When the conical wave has $xz$ as the rotational plane and $x$ as propagating vector, this ``offset force" will be along the $z$-direction. 
This is in good accordance with the inverse Dzyaloshinskii-Moriya effect discussed by Katsura \textit{et al.} \cite{Katsura2005}, Mostovoy \cite{Mostovoy2006c}, and Sergienko \textit{et al.} \cite{Sergienko2006}, who demonstrated that a cycloidal spiral gives rise to a polarization $\vec{P}\propto (\hat{z}\times\hat{e})\times \vec{q}$ with contributions both from electronic charge displacement \cite{Katsura2005,Sergienko2006} and from ionic displacement \cite{Mostovoy2006c,Sergienko2006}. This ferroelectric polarization for cycloidal spirals is unique, since neither a helical spiral nor sinusoidal spin wave states give rise to polarization.



\section{Comparison with expansion of spin exchange parameters}
\label{sec-taylor}

As mentioned before this bilinear formulation of spin-lattice coupling differ from the standard approach.
In the standard formulation
the effective model hamiltonian corresponding to Eq.~(\ref{eq-effHdiscrete}) takes the form
\begin{align}\label{eq-standard}
	\widetilde{\mathcal{H}}_\mathrm{MQ}=-\frac{1}{2}\sum_{ij}\left(
	\bfQ_i\cdot T^{cc}_{ij}\cdot\bfQ_j+
	\bfM_i\cdot \tilde{T}^{ss}_{ij}[\{\bfQ\}]\cdot\bfM_j\right)\,,
\end{align}
where $\tilde{T}^{ss}$ depend on all the ionic displacements $\{\bfQ\}$. Such an expression gives that there is a contribution $\bfF^{sc}_k$ to the total force on site $k$  from an effective spin lattice coupling, 
\begin{align}
 \bfF^{sc}_k= \frac{1}{2}\sum_{ij} \bfM_i\cdot\frac{\partial \tilde{T}^{ss}_{ij}}{\partial\bfQ_k}\cdot\bfM_j\,,\label{eq:TaylorF}
\end{align}
which in general involves a double sum and can be fairly cumbersome to calculate.
However, physically such a derivative can be analyzed in some simple limits. First, in case of pure Heisenberg exchange in nearest neighbor approximation where the isotropic exchange $J$ parameter is dependent on the distance between the two atoms, the exchange tensor can be written as
\begin{align}
	\tilde{T}^{ss}_{ij}[\{\bfQ\}]&=J(|\bfR_{ij}+\bfQ_i-\bfQ_j|)\,{\bf 1}\,,
\end{align}
with the unit tensor $\bf 1$. For such a model the force of Eq.~(\ref{eq:TaylorF}) is only non-vanishing for the two interacting atoms and leads to a derivative 
\begin{align}
	\frac{\partial \tilde{T}^{ss}_{ij}}{\partial\bfQ_i}=-\frac{\partial \tilde{T}^{ss}_{ij}}{\partial\bfQ_j}\approx J'(|\bfR_{ij}|){\bf 1}\,\widehat{\bfR}_{ij}.
\end{align}
The resulting force is in the direction as to gain in Heisenberg exchange energy. Such a force give rise to qualitatively similar results as the present method in the examples of double anti-ferromagnet in \ref{ssec-dafm} and sinusoidal spin density wave in \ref{ssec-planar}.
Second, in the case of the anisotropic Dzyaloshinskii-Moriya interaction $\bfD_\mathrm{DM}$ between two magnetic sites $i$ and $j$ over a bridging ligand site $k$, the interaction vector can in the super-exchange approximation be written as \cite{Sergienko2006}
\begin{align}
\bfD_\mathrm{DM}&\approx D\,\bfR_{ij}\times\bfQ_k\,,
\end{align}
which gives rise to a force on the ligand atom 
\begin{align}
	\bfF^{sc}_k=\frac{1}{2}D\,\bfR_{ij}\times\left(\bfM_i\times\bfM_j\right)\,.
\end{align}
 This result is in qualitative agreement with the present result of the cycloid in \ref{ssec-helical}. In this case $\bfR_{ij}$ lies in the plane spanned by $\bfM_i$ and $\bfM_j$ and a resulting non-oscillating force would be in the same plane but perpendicular to the bond direction, i.e.~what is called $\hat{z}$ in the example above.
 
{To conclude this section we note that in those insulating magnets where the spin texture simultaneously breaks time and spatial reversion, third order spin lattice coupling in Eq.~(\ref{eq:TaylorF}) is commonly considered when describing ferroelectric polarization and multiferroic phases  \cite{Katsura2005,Mostovoy2006c,Sergienko2006,Sergienko2006a}, and also to be responsible for the dynamic magneto-electric response in the electromagnetic field driven dynamics in the GHz and THz regime \cite{Pimenov2006,Sushkov2008,Takahashi2012,Kubacka2014}.} Hopefully, we have here made plausible that the same effects can also be treated in a bi-linear spin-lattice coupling, but a more direct comparison of the two different approaches is left for future studies.
 
 

\section{Equations of motion}
\label{sec-eqm}

\subsection{General dynamical equations}
\label{ssec-general}

Here, we make a brief derivation of the equations of motion that can be obtained from the effective action in Eq. (\ref{eq-effS}). Hence, in order to access the physics in the spin-lattice system we have to convert the time-integration on the Keldysh contour to real times. While all steps in the conversion are shown in Appendix \ref{app-EQM}, we here notice that the transformation leads to a natural introduction of slow and fast spin and lattice variables which, in principle, have to be treated coherently for a complete description. Nevertheless, here we will only address the dynamics of the slow variables in presence of a mean field generated by the fast variables. Accordingly, by differentiating the effective action with respect to the fast variables we can retain a description solely in the slow variables. The conversion to real times does, however, introduce contributions to the model which are quadratic in the fast variables, see Appendix \ref{app-EQM}, such that there remain contributions in the description explicitly depending on these even after differentiating. The simplest solution to this problem is to neglect their existence under the assumption that their overall contribution to the dynamics is negligible. While this approach is somewhat uncontrolled and non-systematic, the equations of motion presented in the main text are obtained in this fashion. A more sophisticated and controlled way to deal with this issue is by application of the Hubbard-Stratonovich transformation, see Appendix \ref{app-quartic}, which leads to that the quadratic terms are replaced by linear ones, however, at the cost of introducing random fields corresponding to quantum fluctuations related to the quadratic spin and lattice interactions.

Here, we adopt the former approach and refer to Appendix \ref{app-quartic} for the details concerning inclusion of the quadratic terms. Our strategy can be justified from the perspective that we here aim to address the general structure of the coupled equations of motion for the spin-lattice system with focus on the contribution that arise from the bi-linear coupling between these subsystems. The resulting equations of motion can be generalized to also include stochastic field of, e.g., Langevin type both addressing the quadratic interaction but also randomness caused by temperature among others. We refer to Appendix \ref{app-quartic} for a discussion of quantum fluctuations caused by rapid spin-spin correlations.

It should also be noticed that through the conversion into real times the interaction fields $\calT_{pq}$ are transformed into retarded/advanced forms, $\calT_{pq}^{r/a}$, which are naturally accessible from electronic structure calculations in terms of the Green functions, see Sec. \ref{sec-derivation}. In this form, we obtain a practical and convenient method to systematically address spin and lattice dynamics at the same level of sophistication and approximation.

Taking the saddle point solution of the total effective spin-lattice action with respect to the fast spin and displacement variables, see Appendix \ref{app-EQM} for details, and requiring $\partial_t|\bfM(x)|^2=0$ for the spin variable, we derive a set of coupled equations of motion given by
\begin{subequations}
\label{eq-eqm}
\begin{align}
\dot\bfM(x)=&
	\bfM(x)
	\times
	\biggl[
		-\gamma\bfB_\text{ext}(x)
\nonumber\\&
		+
		\int
			\Bigl(
				\calT^r_{sc}(x,x')\cdot\bfQ(x')
				+
				\calT^r_{ss}(x,x')\cdot\bfM(x')
			\Bigr)
		dx'
	\biggr]
	,
\label{eq-eqmM}
\\
M_\text{ion}\ddot\bfQ(x)=&
	\gamma_E\bfE_\text{ext}(x)
	+
	\int
		\bfV_{\bfr\bfr'}\cdot\bfQ(x')
		\delta(t-t')
	dx'
\nonumber\\&
		+
		\int
			\Bigl(
				\calT^r_{cs}(x,x')\cdot\bfM(x')
				+
				\calT^r_{cc}(x,x')\cdot\bfQ(x')
			\Bigr)
		dx'
		,
\label{eq-eqmQ}
\end{align}
\end{subequations}
in the presence of external magnetic and electric fields $\bfB_\text{ext}$ and $\bfE_\text{ext}$, respectively. Here, $\dot{\bfM}\equiv\partial_t\bfM$ and $\ddot{\bfQ}\equiv\partial_t^2\bfQ$, whereas the dyad $\bfV_{\bfr\bfr'}\equiv\nabla_\bfr(\nabla_{\bfr'}V_0)$ represents the \emph{ionic contribution to the interatomic force constants}.
The system in Eq. (\ref{eq-eqm}) for $\bfM$ and $\bfQ$ provides a general framework for a coupled treatment of magnetization and lattice dynamics. One should note that Eq. (\ref{eq-eqm}) emphasizes that the temporal and spatial evolution of both $\bfQ$ and $\bfM$ depend non-locally on both the time-dependent magnetization and ionic displacements for the entire structure. The consequence of this non-local description is that all retardation effects within the spin-lattice system that are associated with their coupling to the electronic structure are included in Eq. (\ref{eq-eqm}), despite the seemingly absence of contributions arising from, e.g., damping and moment of inertia \cite{Bhattacharjee2012}. Conceptually, these and other retardation effects are included in the full integration over space and time, however, as we shall see in Sec.~\ref{ssec-adiabatic} it can be shown that damping and moment of inertia are related to temporal expansion of the spin moments. Analogously, the spin-transfer torques can be related to gradient expansion of the magnetization.
In this context it is interesting to observe that the time evolution of a local mode \cite{Hewson2002}, is non-locally influenced by the magnetization at different points in space and time. Due to the coupling it can, moreover, be concluded that the ionic dynamics can be controlled by external magnetic fields, e.g., $\bfB_\text{ext}(x)$, something that was experimentally demonstrated in Ref. \cite{Walker2011}, and reciprocally that magnetic ordering can be driven by electric fields, such as for instance when the electric component of a THz electromagnetic pulse couple to a dipole active phonon mode and excite electromagnons \cite{Pimenov2006,Takahashi2012,Kubacka2014}.

Here it is worth to point out that the uncoupled version of Eq. (\ref{eq-eqmQ}), which describes the ionic vibrations, or, phonons is related to the linear response equations commonly used for such calculations \cite{Baroni2001}. At first glance they look different, but it easy to show that they are closely connected to one formulation of linear response, the so-called dielectric approach \cite{Quong1991,Liu1996}.

The equations of motion presented in Eq. (\ref{eq-eqm}) represent a generalized form of the equations of motion typically used in practical simulations and we will address this issue in Sec. \ref{ssec-adiabatic}. Before entering the next level of approximations, however, it is useful to discuss the general structure of the derived equations.

The first observation one can make is that one retains the uncoupled equations of motion whenever the interaction fields $\calT^r_{sc/cs}\rightarrow0$. In this limit, respective descriptions for lattice and spin dynamics are recovered, however, here provided in a more generalized form since the full retardation (memory) is included in the equations of motion.
Secondly, we notice that the coupling terms $\int\calT_{sc}^r(x,x')\cdot\bfQ(x')dx'$ and $\int\calT_{cs}^r(x,x')\cdot\bfM(x')dx'$ essentially add the effect of an additional magnetic and electric field to the respective equation. These fields are, however, strongly dependent on the properties contained in the interaction tensors $\calT^r_{sc/cs}$ and their couplings to the ionic displacements $\bfQ$ and magnetic moments $\bfM$. The meaning of the statement lies in the fact that these fields may be possible to control through the properties of the electronic structure. In effect, it also leads to that these induced fields can be cancelled or amplified by appropriately choosing and controlling the external electromagnetic fields. Along with the first statement then, this should open for opportunities to make continuous transitions between coupled and uncoupled dynamics by tuning the external fields \cite{Mentink2015}.
As a further implication of this transitioning between the coupled and uncoupled regimes it should become possible to make direct measurements of the frequencies of the uncoupled systems and frequency shifts associated with the coupled dynamics.

\subsection{Adiabatic limit}
\label{ssec-adiabatic}

The temporal non-locality inherited in the equations of motion, Eq. (\ref{eq-eqm}), is of principle value for investigations of the dynamics as it carries the full memory of the time-evolution. In this sense the equations of motion are non-Markovian. Nonetheless, for practical simulations the non-Markovian character presents undesired complications since it requires integrations over all time in addition to keeping track of the full memory of the past at each evaluation of the time-evolution. Moreover, as the equations of motion given in Eq. (\ref{eq-eqm}) are opaque regarding the physical interpretation, the physical meaning of the dynamical exchange interactions $\calT^r_{pq}(x,x')$ is non-trivial to grasp. Therefore, it is meaningful to resort to approximations in the time-domain, if not over all space and time. As we remarked in Sec. \ref{ssec-bilinear}, we shall refer to the adiabatic limit in our discussions of slow temporal and spatial variations of the spin and lattice quantities.

Assuming a slow time-evolution of the spin and displacement variables, we can Taylor expand in the temporal argument to linear order $f(t')\approx f(t)-\tau\dot{f}(t)$, where $\tau=t-t'$. We will, moreover, restrict to the case of small spin fluctuations around a ferromagnetic ground state such that $\dot{\bfM}(\bfr',t)\approx\dot{\bfM}(x)$, as well as slow variations in the displacements such that $\dot{\bfQ}(\bfr',t)\approx\dot{\bfQ}(x)$. Finally, we assume that the interaction tensors have a simple time-dependence, that is, $\calT^r_{pq}(x,x')=\calT^r_{pq}(\bfr,\bfr';t-t')$ which allows to introduce $\calT^r_{pq}(\bfr,\bfr')=\lim_{\omega\rightarrow0}\calT^r_{pq}(\bfr,\bfr';\omega)\equiv\lim_{\omega\rightarrow0}\int\calT^r_{pq}(x,x')e^{i\omega\tau}dt'$. Effecting these assumptions into Eq. (\ref{eq-eqm}), the result can be written as
\begin{subequations}
\label{eq-adiabatic-eqm}
\begin{align}
\dot\bfM(x)=&
	\bfM(x)
	\times
	\biggl(
		-\gamma\bfB(x)
		+
		\hat{G}_{ss}(\bfr)\cdot\dot{\bfM}(x)
		+
		\hat{G}_{sc}(\bfr)\cdot\dot{\bfQ}(x)
	\biggr)
	,
\label{eq-adiabatic-eqmM}
\\
M_\text{ion}\ddot\bfQ(x)=&
	\gamma_E\bfE(x)
	+
	\int
		\bfU_{\bfr\bfr'}\cdot\bfQ(\bfr',t)
	d\bfr'
\nonumber\\&
	+
	\hat{G}_{cc}(\bfr)\cdot\dot{\bfQ}(x)
	+
	\hat{G}_{cs}(\bfr)\cdot\dot{\bfM}(x)
		.
\label{eq-adiabatic-eqmQ}
\end{align}
\end{subequations}

In this set of coupled equations we have introduced the effective magnetic and electric fields $\bfB$ and $\bfE$ which both contain the corresponding external fields and while $\bfB$ also includes both mean fields induced by the surrounding spin and displacement fields, the effective electric field $\bfE$ only additionally includes the mean field of the surrounding spin structure. The effective fields are given by
\begin{subequations}
\label{eq-mfBmfE}
\begin{align}
\bfB(x)=&
	\bfB_\text{ext}(x)
	-
	\frac{1}{\gamma}
	\int
		\Bigl(
			\calT^r_{ss}(\bfr,\bfr')\cdot\bfM(\bfr',t)
\nonumber\\&
			+
			\calT^r_{sc}(\bfr,\bfr')\cdot\bfQ(\bfr',t)
		\Bigr)
	d\bfr'
	,
\label{eq-mfB}
\\
\bfE(x)=&
	\bfE_\text{ext}(x)
	+
	\frac{1}{\gamma_E}
	\int
		\calT^r_{cs}(\bfr,\bfr')\cdot\bfM(\bfr',t)
	d\bfr'
	.
\label{eq-mfE}
\end{align}
\end{subequations}
In this sense the effective magnetic field reduces to the conventional definition in the uncoupled limit while effects of the displacement induced pseudo-magnetic field is included in the coupled regime. Simultaneously, the effective electric field is in the coupled regime modified by the induced electric field from the surrounding spins. Possible displacement induced modifications to the electric field is not included in this contribution. Instead we redefine the ionic contribution to the interatomic force constant to include this field in the expression
\begin{align}
\bfU_{\bfr\bfr'}=&
	\bfV_{\bfr\bfr'}
	+
	\calT^r_{cc}(\bfr,\bfr')
	.
\label{eq-mf}
\end{align}

The dissipative contributions, comprising the rates of change of the spin and displacement variables, can be collected into four the different damping tensors
\begin{subequations}
\label{eq-damping}
\begin{align}
\hat{G}_{pq}(\bfr)=&
	i
	\lim_{\omega\rightarrow0}\partial_\omega
	\int
		\calT^r_{pq}(\bfr,\bfr';\omega)
	d\bfr'
	,\ 
	p,q=s,c
	.
\end{align}
\end{subequations}
The properties of the indirect exchange $\calT^r_{pq}(\bfr,\bfr')$ and damping $\hat{G}_{pq}(\bfr,\bfr')$ can now be discussed in terms of the dynamical interaction $\calT^r_{pq}(\bfr,\bfr';\omega)$ and employing the decoupling introduced the in Sec. \ref{ssec-bilinear}, we can express it as
\begin{align}
\calT^r_{pq}(\bfr,\bfr';\omega)=&
	-
	\int
		\frac{f(\dote{})-f(\dote{}')}{\omega-\dote{}+\dote{}'+i\delta}
		\frac{\Xi_p(\bfr,\bfrho)}{2^{\delta_{ps}}}
		\frac{\Xi_q(\bfrho',\bfr')}{2^{\delta_{qs}}}
\nonumber\\&\hspace{-1cm}\times
		{\rm sp}
			\bfsigma_p
			\im\bfG^r(\bfrho,\bfrho';\dote{})
			\bfsigma_q
			\im\bfG^r(\bfrho',\bfrho;\dote{}')
	\frac{d\dote{}}{2\pi}
	\frac{d\dote{}'}{2\pi}
	d\bfrho
	d\bfrho'
	.
\end{align}
Thus, taking the static limit, $\omega\rightarrow0$, we can write the exchange interaction according to (see Appendix \ref{app-bilinear} for more details) 
\begin{align}
\calT^r_{pq}(\bfr,\bfr')=&
	-\frac{1}{2}
	\im\,{\rm sp}
	\int
		\frac{\Xi_p(\bfr,\bfrho)}{2^{\delta_{ps}}}
		\frac{\Xi_q(\bfrho',\bfr')}{2^{\delta_{qs}}}
\nonumber\\&\times
		f(\dote{})
			\bfsigma_p
			\bfG^r(\bfrho,\bfrho';\dote{})
			\bfsigma_q
			\bfG^r(\bfrho',\bfrho;\dote{})
	\frac{d\dote{}}{2\pi}
	d\bfrho
	d\bfrho'
	.
\end{align}
Analogously, we find the damping tensor given by
\begin{align}
\hat{G}_{pq}(\bfr,\bfr')=&
	-\frac{1}{2}
	{\rm sp}
	\int
		\frac{\Xi_p(\bfr,\bfrho)}{2^{\delta_{ps}}}
		\frac{\Xi_q(\bfrho',\bfr')}{2^{\delta_{qs}}}
\nonumber\\&\times
		f'(\dote{})
			\bfsigma_p
			\im\bfG^r(\bfrho,\bfrho';\dote{})
			\bfsigma_q
			\im\bfG^r(\bfrho',\bfrho;\dote{})
	\frac{d\dote{}}{2\pi}
	d\bfrho
	d\bfrho'
	.
\end{align}
Written in these forms it becomes clear that while the indirect exchange interaction strongly depends both on the structure of the electronic density of states as well as its occupation, Fermi sea property, the properties of the damping is strongly determined by the electronic structure near the Fermi surface, Fermi surface property.

\section{Summary and Conclusions}
\label{sec-summary}

In summary we have constructed a formalism that merges spin and lattice dynamics in a consistent form at the same conceptual level. Starting from a microscopic model of a material, comprising interactions between the delocalized electrons and local magnetic structure, on the one hand, and the lattice distortions, on the other, we derive an effective model which includes the well-known contributions for bi-linear spin-spin and lattice-lattice interactions. The novel aspect of our effective model are contributions that summarize the interactions between the spin and lattice degrees of freedom in a bi-linear form. We, moreover, showed that the interactions are of tensorial nature which preserve time-reversal and inversion symmetries between the spin and lattice subsystems.

Our findings provide a fundamental new and novel perspective in the theoretical modelling of coupled spin and lattice reservoirs for both, dynamical and static properties. For this purpose, multiple achievements were put into practise: \textit{i)} both spin and lattice reservoirs are treated on the same footing by means of local couplings of the electronic structure with the magnetization on one hand and with lattice distortions on the other hand. These local couplings lead to an effective electron mediated spin-lattice coupling. Such type of spin-lattice-coupling was obtained from the effective action of the system, shown not to violate fundamental symmetry operations of the total energy. Couplings of this nature were, moreover, numerically determined and analytically corroborated from model electronic structure theory for certain magnetic textures, that exists in nature and are already catalogued \cite{Fawcett1988,Sjostedt2002,Walser2012,Fust2016}. On the sidelines, a Green function formalism for pure spin-spin and lattice-lattice second-order rank couplings in agreement with already established methods \cite{Udvardi2003,Diaz-Sanchez2007} was realised.
\textit{ii)} The derived equations of motion account for the most general dynamics of the coupled spin-lattice reservoir, including space-time retardation that causes, for instance, energy dissipation through the Gilbert damping \cite{Kambersky1984,Gilmore07} as well as higher order conservative forces as the moment of inertia \cite{Bhattacharjee2012,Thonig2017}. In principle, also thermal microscopic fields beyond the white-noise and Markovian ansatz \cite{Risken1989tfe,Hellsvik10}, due to the fluctuation-dissipation theorem, are considered. The Gilbert damping $\hat{G}^{ss}$, given in terms of multiple scattering was provided in Ref.~\cite{Bhattacharjee2012}, but the corresponding ionic displacement damping $\hat{G}^{cc}$ and the mixed spin-lattice damping tensors $\hat{G}^{cs}$ and $\hat{G}^{sc}$, are provided as generalizations of these expressions.

The proposed analytical formalism and first numerical results encourage for more detailed theoretical studies. In particular, it motivates to include bilinear spin-lattice coupling in combined classical atomistic spin-lattice dynamics \cite{Ma2008,Ma2012a}, but also to account for exact energy dissipations caused space-time retardation in the equation of motion. All proposed terms $\{T^{pq}\}$ and $\{\hat{G}^{pq}\}, p,q=s,c$, can be implemented in first principles calculations in a similar manner as for the magnetic exchange interactions, which is nowadays a standard tool in various codes. A detailed materials-specific characterization of bilinear spin-lattice couplings is necessary to propose classes of materials with large $\{T^{cs}\}$. The strong hybridisation of the spin and lattice quasiparticle spectra caused by this type of coupling and, thus, possibly enhanced group velocities of the quasi-particles could lead to significant improvements in magnonics and phononics applications. 

In particular, finite temperature phenomena pertaining to the bilinear spin-lattice coupling are highly interesting in, for instance, how critical indices of magnetic or ferroelectric phase transitions change or how phonon and spin temperatures in terms of disorder in the system are affected.

Our study requests also novel experiments, as for instance neutron scattering measurements, to approve the existence of a bilinear spin-lattice coupling in the here proposed magnetic textures. 
Within the formalism, higher order interactions, as three and four body interactions including lattice anharmonicity, are accessible in a systematic way, something which would be of great value for deeper investigations of non-equilibrium dynamics on ultrafast time-scales.

\acknowledgements
We further thank A. V. Balatsky, A. Bergman, J. Lorenzana, P. W\"olfle, and J.-X. Zhu for valuable comments. This work was supported by the Vetenskapsr\aa det, the Wenner-Gren foundation, the Icelandic Research Fund (Grant No. 163048-052), the mega-grant of the Ministry of Education and Science of the Russian Federation (grant no. 14.Y26.31.0015), and Stiftelsen Olle Engqvist Byggm\"astare. J.F. gratefully acknowledges the generous hospitality shown by the T-Division at Los Alamos National Laboratory during his stay in 2012. J.H. is partly funded by the Swedish Research Council (VR) through a neutron project grant (BIFROST, Dnr. 2016-06955).

\appendix
\section{Derivation of effective spin-lattice model}
\label{app-derivation}

\label{app-spin-lattice}
Throughout Secs. \ref{app-derivation} -- \ref{app-quartic} the notation will refer to quantities that are continuous in the spatial dimensions, that is, $A=A(\bfr)$, where $\bfr$ denotes the spatial coordinates. While this is made for mathematical convenience it is straight forward to reduce to discrete lattice structures by defining the quantity $A$ on the lattice through $A(\bfr)=\sum_mA\delta(\bfr-\bfr_m)$, where $\bfr_m$ denotes the lattice coordinate.

\subsection{Microscopic model}
\label{aapp-microscopic}
We model the magnetic interactions by assuming that the magnetization $\bfM(\bfr)$ interacts with the surrounding spin density $\bfs_s(\bfr)$ via the interaction Hamiltonian
\begin{align}
\Hamil_M=&
	-\int v(\bfr,\bfr')\bfM(\bfr)\cdot\bfs_s(\bfr')d\bfr d\bfr'
	.
\end{align}
Here, $\bfs_s(\bfr)\equiv\psi^\dagger(\bfr)\bfsigma\psi(\bfr)/2$ is defined in terms of the spinor $\psi(\bfr)=(\psi_\up(\bfr)\ \psi_\down(\bfr))^T$, whereas $v(\bfr,\bfr')=v(\bfr',\bfr)$ corresponds to the direct exchange contribution from the Coulomb integral.
 
The charge $n(\bfr)=\bfs_c(\bfr)\equiv\psi^\dagger(\bfr)\psi(\bfr)$ is subject to the potential $\phi(\bfr)=\int\phi(\bfr,\bfQ(\bfr'))d\bfr'$ due to electron-ion interactions, where $\bfQ(\bfr)$ is the ionic displacement from its equilibrium position. Here, we do not assign any specific nature of the displacement. For small displacements, we employ the expansion $\phi(\bfr,\bfQ(\bfr'))\approx\phi_0(\bfr)+\bfQ(\bfr')\cdot\nabla_{\bfr'}\phi_0(\bfr)$, where $\phi_0(\bfr)=\lim_{\bfQ\rightarrow0}\phi(\bfr)$, which gives the interaction between the charge and lattice vibrations
\begin{align}
\Hamil_\text{ep}=&
	\int\bfQ(\bfr')\cdot\bflambda(\bfr,\bfr')n(\bfr)d\bfr d\bfr'
	,
\end{align}
where the electron-phonon coupling is denoted by $\bflambda(\bfr,\bfr')=\lim_{\bfr'\rightarrow\bfr}\nabla_{\bfr'}\phi_0(\bfr')$.

\subsection{Effective action}
\label{aapp-microscopic}
Given the general non-equilibrium conditions in the system, e.g. temporal fluctuations and currents, we define the corresponding action on the Keldysh contour \cite{Eckern1984,Larkin1983,Zhu2004,Fransson2008b,Fransson2010,Fransson2014,Bhattacharjee2012} according to
\begin{align}
\calS=&
	\int(\Hamil_\text{M}+\Hamil_\text{ep})dt
	+\calS_B
	+\calS_\text{WZWN}
	+\calS_\text{latt}
	+\calS_E
	.
\end{align}
Here,
\begin{align}
\calS_B=&
	-\gamma\int\bfB_\text{ext}(x)\cdot\bfM(x)dx
	,
\end{align}
$x=(\bfr,t)$, describes the Zeeman coupling to the external magnetic field $\bfB_\text{ext}(x)$, whereas
\begin{align}
\calS_{WZWN}=&
	\int\int_0^1
		\frac{\bfM(x;\tau)}{|\bfM(\bfr)|^2}
		\cdot
		\Bigl[
			\partial_\tau\bfM(x;\tau)
			\times
			\partial_t\bfM(x;\tau)
		\Bigr]
	d\tau
	dx
\end{align}
accounts for the Berry phase accumulated by the spin. The free lattice is represented by, e.g.,
\begin{align}
\calS_\text{latt}=&
	\int
		\biggl\{
			\biggl[
				i\bfQ(x)
				\cdot
				\dt\bfQ(x)
				-
				\frac{M_\ion}{2}
				\{\partial_t\bfQ(x)\}^2
			\biggr]
			\delta(\bfr-\bfr')
\nonumber\\&
			-
			\bfQ(x)
			\cdot
			\bfV_{\bfr\bfr'}
			\cdot
			\bfQ(\bfr',t)
		\biggr\}
	d\bfr'
	dx
	,
\end{align}
with the ionic mass $M_\ion$ and the dyad $\bfV_{\bfr\bfr'}=\nabla_{\bfr}(\nabla_{\bfr'}V_0)$ is the \emph{ionic contribution to the interatomic force constant}, and where $V_0$ is the ionic potential at equilibrium (vanishing displacements). Finally, the coupling between the lattice and the external electric field $\bfE_\text{ext}(x)$ is given by
\begin{align}
\calS_E=&
	\int\gamma_E(x)\bfE_\text{ext}(x)\cdot\bfQ(x)dx
	,
\end{align}
where $\gamma_E(x)$ essentially comprises the displaced charge at $x$.

We obtain an effective action $\calS_\text{MQ}$ for the coupled magnetization and lattice dynamics through a second order cumulant expansion of the partition function $\calZ[\bfM(x),\bfQ(x)]={\rm Tr\, T_C}e^{i\calS}$ and tracing over the electronic degrees of freedom (${\rm Tr}$). The result can be written
\begin{align}
\calS_\text{MQ}=&
	-\frac{1}{2}\int
	\Bigl(
			\bfQ(x)\cdot
			[
				\calT_{cc}(x,x')\cdot\bfQ(x')
				+\calT_{cs}(x,x')\cdot\bfM(x')
			]
\nonumber\\&
			+\bfM(x)\cdot
			[
				\calT_{sc}(x,x')\cdot\bfQ(x') 
				+\calT_{ss}(x,x')\cdot\bfM(x')
			]
	\Bigr)
	dxdx',
\end{align} 
where we have defined the interaction tensor
\begin{subequations} 
\begin{align} 
\calT_{pq}(x,x')=& 
	\int
		\Xi_p(\bfr,\bfrho)
		\bfK_{pq}(\bfrho,\bfrho';t,t)
		\Xi_q(\bfrho',\bfr')
	d\bfrho d\bfrho',
\label{eq-Dpq}
\\
\bfK_{pq}(\bfrho,\bfrho';t,t')=&
	\eqgr{\bfs_p(\bfrho,t)}{\bfs_q(\bfrho',t')},
\label{eq-Kpq}
\end{align}
\end{subequations}
with the notation $\Xi_c(\bfr,\bfrho)=\nabla_\bfr\phi_c(\bfrho)$, and $\Xi_s(\bfr,\bfrho)=-v(\bfr,\bfrho)$, for $p,q=c,s$.

\section{Equations of motion}
\label{app-EQM}

\label{app-eqm}
The time integrals in Eq. (\ref{eq-effS}) run over the (Keldysh) contour, $C$, in the complex plane and have to be converted into real time integrals. This can be done by the following procedure. The contour $C$ has one branch above and one below the real axis, and we therefore label the involved variables $\bfM$ and $\bfQ$ with superscripts $u$ and $l$ for the upper and lower branches, respectively. Likewise, we introduce the real time ordered and anti-time ordered propagators $\calT_{pq}^{t/\bar{t}}(x,x')$ for times $t,\ t'$ both on the upper/lower branch and  $\calT_{pq}^{</>}(x,x')$ for $t$ on the upper/lower branch and $t'$ on the lower/upper. Using this notation we have, for instance,
\begin{align}
\int\bfM(x)\cdot\calT_{ss}(x,x')\cdot\bfM(x')dxdx'&
\nonumber\\&\hspace{-4cm}=
	\int_{-\infty}^\infty
	\begin{pmatrix}
		\bfM^u(x) &
		-\bfM^l(x)
	\end{pmatrix}
\nonumber\\
	\cdot
	\begin{pmatrix}
		\calT_{ss}^t(x,x') & \calT_{ss}^<(x,x') \\
		\calT_{ss}^>(x,x') & \calT_{ss}^{\bar{t}}(x,x')
	\end{pmatrix}
	&
	\cdot
	\begin{pmatrix}
		\bfM^u(x') \\
		-\bfM^l(x')
	\end{pmatrix}
	dxdx'.
\end{align}
Here, the dot ($\cdot$) between the matrices is retained as a reminder that each contribution to this expression is composed of a product of the type $\bfM\cdot\calT\cdot\bfM$.
Using the unitary rotation
\begin{align}
\calR=&
	\frac{1}{\sqrt{2}}
	\begin{pmatrix}
		1 & 1 \\ 
		-1 & 1
	\end{pmatrix},
\end{align}
the above expression becomes
\begin{align}
\frac{1}{2}
	\int&
	\begin{pmatrix}
		2\bfM^c(x) &
		\bfM^q(x)
	\end{pmatrix}
\nonumber\\&
	\cdot
	\begin{pmatrix}
		0 & \calT_{ss}^a(x,x') \\
		\calT_{ss}^r(x,x') & \calT_{ss}^K(x,x')
	\end{pmatrix}
	\cdot
	\begin{pmatrix}
		2\bfM^c(x') \\
		\bfM^q(x')
	\end{pmatrix}
	dxdx',
\end{align}
where we have introduced new (slow/fast) variables $\bfM^c\equiv(\bfM^u+\bfM^l)/2$ and $\bfM^q=\bfM^u-\bfM^l$, requiring $\bfM^c\cdot\bfM^q=0$, and the retarded/advanced/Keldysh propagators $\calT^{r/a/K}_{ss}$ with
\begin{subequations}
\begin{align}
\bfK_{ss}^{r/a}(\bfrho,\bfrho';t,t')=&
	(\mp i)\theta(\pm t-\mp t')\av{\com{\bfs(\bfrho,t)}{\bfs(\bfrho',t')}},
\label{eq-Kra}
\\
\bfK_{ss}^K(\bfrho,\bfrho';t,t')=&
	(-i)\av{\anticom{\bfs(\bfrho,t)}{\bfs(\bfrho',t')}}.
\label{eq-KK}
\end{align}
\end{subequations}
Here, the brackets, Eq. (\ref{eq-Kra}), and braces, Eq. (\ref{eq-KK}), refer to commutation and anti-commutation, respectively.
Noticing that $\int\bfM^c(x)\cdot\calT_{ss}^a(x,x')\cdot\bfM^q(x')dxdx'=\int\bfM^q(x)\cdot\calT_{ss}^r(x,x')\cdot\bfM^c(x')dxdx'$, we can write
\begin{align}
2\int&
	\bfM^q(x)\cdot
	\left[
		\calT^r_{ss}(x,x')\cdot\bfM^c(x')
		+\frac{1}{4}\calT^K_{ss}(x,x')\cdot\bfM^q(x')
	\right]
	dxdx'.
\end{align}
In this fashion, we obtain one contribution which is linear, and one which is quadratic, in the fast variables. For now, we will omit the quadratic contributions. In App. \ref{app-quartic} we will return to this issue and show how those quadratic contributions can be related to quantum (spin-spin, lattice-lattice, and spin-lattice) fluctuations and included in the formalism through introduction of random variables.

The equations of motion for the magnetization $\bfM$ and displacement $\bfQ$ are found by variation of the total action $\calS$ with respect to fast fluctuations, see e.g. Ref. \cite{Bhattacharjee2012} for details. Requiring $\partial_t|\bfM(x)|^2=0$, we obtain
\begin{subequations}
\label{aeq-eqm}
\begin{align} 
\dot\bfM(x) 
	=& 
	\bfM(x)\times
	\biggl(
		-\gamma\bfB_\text{ext}(x) 
\nonumber\\& 
		+\int 
		\Bigl( 
			\calT_{sc}^r(x,x')\cdot\bfQ(x') 
			+\calT_{ss}^r(x,x')\cdot\bfM(x') 
		\Bigr)
		dx'
	\biggr),
\label{aeq-eqmM}
\\ 
M_\ion\ddot\bfQ(x)=&
	\gamma_E(x)\bfE_\text{ext}(x)
	+
	\int
		\bfV_{\bfr\bfr'}\cdot\bfQ(\bfr',t)
	d\bfr'
\nonumber\\&
	+
	\int 
	\biggl( 
		\calT_{cs}^r(x,x')\cdot\bfM(x')
		+\calT_{cc}^r(x,x')\cdot\bfQ(x') 
	\biggr) 
	dx'
	,
\label{aeq-eqmQ}
\end{align} 
\end{subequations}
where the superscript $r$ refers to retarded propagators.

\section{Quantum Fluctuations}
\label{app-quartic}

The expansion of the action on the Keldysh contour leads to contributions which are quadratic in the superscript $q$, and have thus been omitted so far. Here, we shall study the effect of those contributions by Bosonization as accomplished through the Hubbard-Stratonovich transformation.

Following Ref. \cite{Negele88} we notice that e.g. the contribution
\begin{align}
e&^{-\frac{i}{4}\int\bfM^q(x)\cdot\calT^K(x,x')\cdot\bfM^q(x')dx dx'}
\nonumber\\&
=
	\int\calT\bfxi
		e^{\frac{i}{4}\int\bfxi(x)\cdot\calT^{K,-1}(x,x')\cdot\bfxi(x')dxdx'}
		e^{-\frac{i}{2}\int\bfxi(x)\cdot\bfM^q(x)dx},
\label{eq-HS}
\end{align}
where the measure $\calT\bfxi=\lim_{\epsilon\rightarrow0}\prod\sqrt{\det{\dote{}\calT^{K,-1}}/i2\pi}d\bfxi$, whereas the random fields $\bfxi$ can be related to the spin-susceptibility $\calT^K$ through the following procedure. In Eq. (\ref{eq-HS}), by $\calT^{K,-1}$ we mean the inverse of $\calT^K$. Assuming that there is a random magnetic field $\bfxi$ coupled to the magnetization variable $\bfM^q$ through $\Hamil_\xi=-\gamma_\xi\bfxi\cdot\bfM^q$. Then, with respect to these random fields, the partition function can be written
\begin{align}
\calZ[\bfxi]=
	\tr_\bfxi e^{-\int\Hamil_\xi(t)dt}
	\approx&
	e^{-\frac{1}{2}\int\av{\Hamil_\xi(t)\Hamil_\xi(t')}dtdt'}
\nonumber\\=&
	e^{-\gamma_\xi^2\int\bfM^q(x)\cdot\av{\bfxi(x)\bfxi(x')}\cdot\bfM^q(x')dxdx'/2}
	.
\end{align}
Inspection of the two equations suggests that the random variables $\bfxi$ have to satisfy the condition
\begin{align}
\av{\bfxi(x)\bfxi(x')}=&
	\frac{i}{2\gamma_\xi^2}\calT^K(x,x').
\end{align}
Recall that $\calT^K(x,x')$ denotes the Keldysh field defined in terms of the kernel in Eq. (\ref{eq-KK}). We also remark that this relation is a clear manifestation that the quantum correlation induced noise is not necessarily of white Gaussian nature. It also shows that the quantum noise strongly depends on the electronic structure.
We can generalize this procedure to the whole action $\calS_q$ since we can write (omitting the superscripts $q$ and $K$)
\begin{align}
e^{i\calS_q}=&
	\exp\left\{
		-\frac{i}{4}
		\int
		\begin{pmatrix} \bfM & \bfQ \end{pmatrix}
		\cdot
		\begin{pmatrix} 
			\calT_{ss} & \calT_{sc} \\
			\calT_{cs} & \calT_{cc}
		\end{pmatrix}
		\cdot
		\begin{pmatrix}
			\bfM \\
			\bfQ
		\end{pmatrix}
		d\mu(x,x')
		\right\}
\nonumber\\=&
	\exp\left\{-\frac{i}{4}a_iA_{ij}a_j\right\},
\end{align}
where $a_1=\bfM(x)$ and $a_2=\bfQ(x)$. By means of the Hubbard-Stratonovich transformation we now obtain
\begin{align}
e&^{i\calS_q}
\nonumber\\=&
	\int\calT\bfphi
		e^{\frac{i}{4}\int\bfphi(x)A^{-1}(x,x')\bfphi(x')d\mu(x,x')}
		e^{-\frac{i}{2}\int\bfphi(x)\cdot\bfa(x)d\mu(x)}.
\end{align}
Defining the random variables $\bfxi$ and $\bfzeta$ such that $\bfphi\cdot\bfa=\bfxi\cdot\bfM+\bfzeta\cdot\bfQ$, we can relate those random variables to the correlation functions $\calT_{pq}$, through
\begin{align}
	\av{\bfphi(x)\bfphi^T(x')}
	=&
	\frac{i}{2}
	\begin{pmatrix}
		\frac{1}{\gamma_\xi} & \frac{1}{\gamma_\zeta}
	\end{pmatrix}
	\begin{pmatrix}
		\calT_{ss}^K(x,x') & \calT_{sc}^K(x,x') \\
		\calT_{cs}^K(x,x') & \calT_{cc}^K(x,x')
	\end{pmatrix}
	\begin{pmatrix}
		\frac{1}{\gamma_\xi} \\ \frac{1}{\gamma_\zeta}
	\end{pmatrix}
	,
\end{align}
where $\bfphi(x)=\begin{pmatrix}\bfxi(x) & \bfzeta(x)\end{pmatrix}^T$.

The contribution to the spin-lattice coupled system can, thus, be written as
\begin{align}
\calS_q=&
	-\frac{1}{2}\int
		\Bigl(
			\gamma_\xi\bfxi\cdot\bfM^q(x)
			+\gamma_\zeta\bfzeta\cdot\bfQ^q(x)
		\Bigr)
		d\bfr dt.
\end{align}
This action, which is due to the fast correlations between the magnetization and lattice dynamics, adds correlation effects to the equations of motion for $\bfM$ and $\bfQ$ via the random fields $\bfxi$ and $\bfzeta$. The random fields $\bfxi$ and $\bfzeta$ relates to the spin-spin, lattice-lattice, and spin-lattice interactions via their corresponding correlation function. Those new Bosonic degrees of freedom represent collective modes that are associated with fluctuations in the magnetic and lattice structure, that is, spin waves (magnons) and lattice vibrations (phonons).

\section{Inversion symmetry}
\label{App-Symm}

\begin{figure}
\centering
 \includegraphics[scale=0.3]{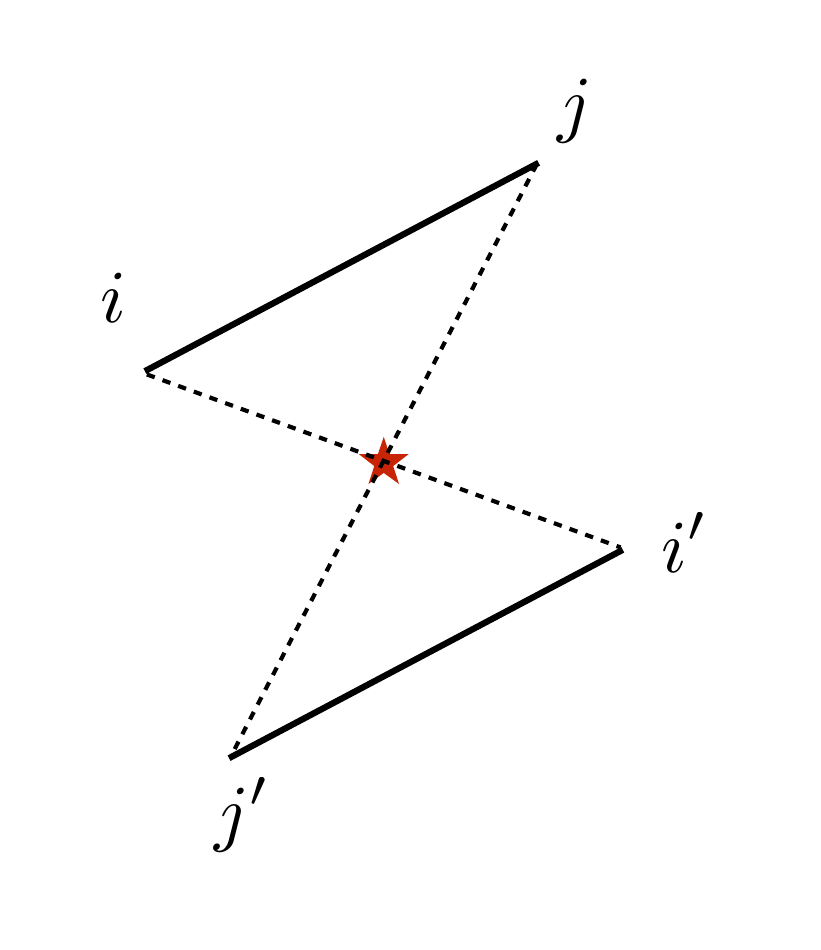}
\caption{Interactions (full lines) between two sites $i$ and $j$ that are connected (dashed lines) to sites $i'$ and $j'$ by inversion. }
\label{fig:Inversion}
\end{figure}

If inversion $\iota$ is a symmetry operation that bring site $i$ to site $i'$ as in Figure \ref{fig:Inversion}, we can focus
 on the two pair interactions $ij$ respectively $i'j'$. 
In order to study these in detail we introduce the average quantities
\begin{align}
	\vec{Q}&=\vec{Q}_i+\vec{Q}_j+\vec{Q}_{i'}+\vec{Q}_{j'}=\vec{Q}_{ii'}+\vec{Q}_{jj'}\nonumber\\
	\vec{q}_1&=\vec{Q}_i-\vec{Q}_j+\vec{Q}_{i'}-\vec{Q}_{j'}=\vec{Q}_{ii'}-\vec{Q}_{jj'}\nonumber\\
	\vec{q}_2&=\vec{Q}_i+\vec{Q}_j-\vec{Q}_{i'}-\vec{Q}_{j'}=\vec{q}_{ii'}+\vec{q}_{jj'}\nonumber\\
	\vec{q}_3&=\vec{Q}_i-\vec{Q}_j-\vec{Q}_{i'}+\vec{Q}_{j'}=\vec{q}_{ii'}-\vec{q}_{jj'}\,,
	\label{Eq:Qaver}
\end{align}
and
\begin{align}
	\vec{M}&=\vec{M}_i+\vec{M}_j+\vec{M}_{i'}+\vec{M}_{j'}=\vec{M}_{ii'}+\vec{M}_{jj'}\nonumber\\
    \vec{m}_1&=\vec{M}_i-\vec{M}_j+\vec{M}_{i'}-\vec{M}_{j'}=\vec{M}_{ii'}-\vec{M}_{jj'}\nonumber\\
        \vec{m}_2&=\vec{M}_i+\vec{M}_j-\vec{M}_{i'}-\vec{M}_{j'}=\vec{m}_{ii'}+\vec{m}_{jj'}\nonumber\\
            \vec{m}_3&=\vec{M}_i-\vec{M}_j-\vec{M}_{i'}+\vec{M}_{j'}=\vec{m}_{ii'}-\vec{m}_{jj'}\,,\label{Eq:Maver}
\end{align}
that all  have well defined parity properties
\begin{align}
	\iota\vec{Q}&=-\vec{Q}_{ii'}-\vec{Q}_{jj'}=-\vec{Q}\nonumber\\
		\iota\vec{q}_1&=-\vec{Q}_{ii'}+\vec{Q}_{jj'}=-\vec{q}_1\nonumber\\
		\iota\vec{q}_2&=-\vec{q}_{ii'}+\vec{q}_{jj'}=\vec{q}_2\nonumber\\
		\iota\vec{q}_3&=\vec{q}_{ii'}-\vec{q}_{jj'}=\vec{q}_3\,,
\end{align}
and
\begin{align}
			\iota\vec{M}&=\vec{M}_{ii'}+\vec{M}_{jj'}=\vec{M}\nonumber\\
\iota\vec{m}_1&=\vec{M}_{ii'}-\vec{M}_{jj'}=\vec{m}_1\nonumber\\
\iota\vec{m}_2&=-\vec{m}_{ii'}-\vec{m}_{jj'}=-\vec{m}_2\nonumber\\
\iota\vec{m}_3&=-\vec{m}_{ii'}+\vec{m}_{jj'}=-\vec{m}_3\,.
\end{align}
Then since the individual quantities can be obtained by reversing the Eqs.~(\ref{Eq:Qaver}) and (\ref{Eq:Maver})
\begin{align}
\vec{Q}_i&=\frac{1}{4}\left(\vec{Q}+\vec{q}_1+\vec{q}_2+\vec{q}_3\right)\nonumber\\
\vec{Q}_j&=\frac{1}{4}\left(\vec{Q}-\vec{q}_1+\vec{q}_2-\vec{q}_3\right)\nonumber\\
\vec{Q}_{i'}&=\frac{1}{4}\left(\vec{Q}+\vec{q}_1-\vec{q}_2-\vec{q}_3\right)\nonumber\\
\vec{Q}_{j'}&=\frac{1}{4}\left(\vec{Q}-\vec{q}_1-\vec{q}_2+\vec{q}_3\right)\,,
\end{align}
and
\begin{align}
\vec{M}_i&=\frac{1}{4}\left(\vec{M}+\vec{m}_1+\vec{m}_2+\vec{m}_3\right)\nonumber\\
\vec{M}_j&=\frac{1}{4}\left(\vec{M}-\vec{m}_1+\vec{m}_2-\vec{m}_3\right)\nonumber\\
\vec{M}_{i'}&=\frac{1}{4}\left(\vec{M}+\vec{m}_1-\vec{m}_2-\vec{m}_3\right)\nonumber\\
\vec{M}_{j'}&=\frac{1}{4}\left(\vec{M}-\vec{m}_1-\vec{m}_2+\vec{m}_3\right)
\end{align}
\begin{widetext}
we can rewrite the  pair tensors interactions between sites $i$ and $j$  of Eq.~(\ref{pairinter}) as
\begin{align}
	I_{ij}=\,
	&\mathcal{S}_{ij}\left( \vec{Q}_i\vec{M}_j+\vec{Q}_j\vec{M}_i\right) 
		+\mathcal{A}_{ij}\left( \vec{Q}_i\vec{M}_j-\vec{Q}_j\vec{M}_i\right)
		=\nonumber\\
	=&\frac{1}{8}\mathcal{S}_{ij}
	\big\{ \vec{Q} \vec{M}+\vec{Q} \vec{m}_2-\vec{q}_1 \vec{m}_1-\vec{q}_1 \vec{m}_3+\vec{q}_2 \vec{M}+\vec{q}_2 \vec{m}_2-\vec{q}_3 \vec{m}_1-\vec{q}_3 \vec{m}_3\big\}+\nonumber\\
	+&\frac{1}{8}\mathcal{A}_{ij}\big\{-\vec{Q} \vec{m}_1-\vec{Q} \vec{m}_3+\vec{q}_1 \vec{M}+\vec{q}_1 \vec{m}_2-\vec{q}_2 \vec{m}_1-\vec{q}_2 \vec{m}_3+\vec{q}_3 \vec{M}+\vec{q}_3 \vec{m}_2\big\}\,,
\end{align}
while the corresponding interactions between $i'$ and $j'$ become
\begin{align}
I_{i'j'}=\,
	&\mathcal{S}_{i'j'} \left( \vec{Q}_{i'}\vec{M}_{j'}+\vec{Q}_{j'}\vec{M}_{i'}\right)
		+\mathcal{A}_{i'j'}\left( \vec{Q}_{i'}\vec{M}_{j'}-\vec{Q}_{j'}\vec{M}_{i'}\right)
		=\nonumber\\
	=&\frac{1}{8}\mathcal{S}_{i'j'}
	\big\{ \vec{Q} \vec{M}-\vec{Q} \vec{m}_2-\vec{q}_1 \vec{m}_1+\vec{q}_1 \vec{m}_3-\vec{q}_2 \vec{M}+\vec{q}_2 \vec{m}_2+\vec{q}_3 \vec{m}_1-\vec{q}_3 \vec{m}_3\big\}+\nonumber\\
	+&\frac{1}{8}\mathcal{A}_{i'j'} 
\big\{ -\vec{Q} \vec{m}_1+\vec{Q} \vec{m}_3+\vec{q}_1 \vec{M}-\vec{q}_1 \vec{m}_2+\vec{q}_2 \vec{m}_1-\vec{q}_2 \vec{m}_3-\vec{q}_3 \vec{M}+\vec{q}_3 \vec{m}_2    \big\}\,.
\end{align}
As
\begin{align}
	\iota &\big\{\vec{Q} \vec{M}+\vec{Q} \vec{m}_2-\vec{q}_1 \vec{m}_1-\vec{q}_1 \vec{m}_3+\vec{q}_2 \vec{M}+\vec{q}_2 \vec{m}_2-\vec{q}_3 \vec{m}_1-\vec{q}_3 \vec{m}_3\big\}=\nonumber\\
	=&-\big\{ \vec{Q} \vec{M}-\vec{Q} \vec{m}_2-\vec{q}_1 \vec{m}_1+\vec{q}_1 \vec{m}_3-\vec{q}_2 \vec{M}+\vec{q}_2 \vec{m}_2+\vec{q}_3 \vec{m}_1-\vec{q}_3 \vec{m}_3\big\}\nonumber\\
	\iota & \big\{ -\vec{Q} \vec{m}_1+\vec{Q} \vec{m}_3+\vec{q}_1 \vec{M}-\vec{q}_1 \vec{m}_2+\vec{q}_2 \vec{m}_1-\vec{q}_2 \vec{m}_3-\vec{q}_3 \vec{M}+\vec{q}_3 \vec{m}_2    \big\}=\nonumber \\
	=& -\big\{ -\vec{Q} \vec{m}_1+\vec{Q} \vec{m}_3+\vec{q}_1 \vec{M}-\vec{q}_1 \vec{m}_2+\vec{q}_2 \vec{m}_1-\vec{q}_2 \vec{m}_3-\vec{q}_3 \vec{M}+\vec{q}_3 \vec{m}_2    \big\}\,,
\end{align}
in order to preserve the inversion symmetry, i.e.~that $\iota I_{ij}=I_{i'j'}$, we can identify that both interaction parameters have to have odd parity
\begin{align}
	\iota \mathcal{S}_{ij}&=-\mathcal{S}_{i'j'}\nonumber\\
	\iota \mathcal{A}_{ij}&=-\mathcal{A}_{i'j'}\,.
\end{align}

\end{widetext}

\section{Examples}
\label{app-Examples}

We assume a simple two-dimensional electrons gas, for example, surface states on a metallic surface or an analogous set-up, in which magnetic defects are embedded. We model this system by the Hamiltonian
\begin{align}
\Hamil=&
	\sum_\bfk\Psi_\bfk^\dagger\bfepsilon_\bfk\Psi_\bfk
	+
	\int
		\Psi^\dagger(\bfr)\bfV(\bfr)\Psi(\bfr)
	d\bfr
	,
\end{align}
where the spinor $\Psi_\bfk=(\cs{\bfk\up}\ \cs{\bfk\down})^t$ annihilates electrons with energy $\bfepsilon_\bfk=\dote{\bfk}\sigma^0$ at the momentum $\bfk$ and spin $\sigma=\up,\down$, whereas the scattering potential $\bfV(\bfr)=\sum_m\bfV_m\delta(\bfr-\bfr_m)$ defines a collection of defects $\bfV_m=V_m\sigma^0+\bfM_m\cdot\bfsigma$.

In this model, the unperturbed Green function $\bfg_\bfk$ is defined for the first term and is given in reciprocal and real space by the expressions
\begin{align}
\bfg_\bfk(\omega)=
	\frac{\sigma^0}{\omega-\dote{\bfk}+i\delta}
	,
	&&
\bfg(\bfr,\omega)=&
	-i\frac{N_0}{2}H_0^{(1)}(\kappa r)\sigma^0
	,
\end{align}
where $\kappa^2=2N_0\omega$ and $N_0=m_e/\hbar^2$, whereas $H_m^{(1)}$ is the Hankel function of first kind and order $m$, and $m_e$ is the effective electron mass. In this way we have defined $\bfg_\bfk(\omega)=g_0(\bfk,\omega)\sigma^0$ while $\bfg_1(\bfk,\omega)\equiv0$.

We calculate the dressed Green function $\bfG$ in terms of the $T$-matrix expansion of the impurity potential, that is, 
\begin{align}
\bfG(\bfk,\bfk')=&
	\delta(\bfk-\bfk')\bfg_\bfk
	+
	\sum_{mn}
		\bfg_\bfk
		e^{-i\bfk\cdot\bfr_m}
		\bfT(\bfR_{mn})e^{i\bfk'\cdot\bfr_n}
		\bfg_{\bfk'}
	,
\end{align}
where $\bfR_{mn}=\bfr_m-\bfr_n$, whereas the $T$-matrix is given by
\begin{subequations}
\begin{align}
\bfT(\bfR_{mn})=&
	\bfV_m(\bft^{-1})_{mn}
	,
\\
\bft_{mn}=&
	\delta_{mn}\sigma^0
	+
	\bfg(\bfR_{mn})\bfV_n
	.
\end{align}
\end{subequations}
Here, since the scattering potential is partitioned into a non-magnetic and a magnetic component, we can write $\bfT(\bfR_{mn})=T_0(\bfR_{mn})\sigma^0+\bfT_1(\bfR_{mn})\cdot\bfsigma$.

For sufficiently large separation between the scattering impurities, the $T$-matrix reduces to
\begin{subequations}
\begin{align}
\bfT(\bfR_{mn})=&
	\delta(\bfR_{mn})
	\Bigl[
		\bfV_m^{-1}-\bfg(\bfr=0)
	\Bigr]^{-1}
\nonumber\\=&
	\delta(\bfR_{mn})
	\Bigl[
		t_0(\bfr_m)\sigma^0+\bft_1(\bfr_m)\cdot\bfsigma
	\Bigr]^{-1}
	,
\\
t_0(\bfr_m)=&
	\frac{V_m+i(V_m^2-|\bfM_m|^2)N_0/2}
		{1-(V_m^2-|\bfM_m|^2)(N_0/2)^2+iV_mN_0}
	,
\\
\bft_1(\bfr_m)=&
	\frac{\bfM_m}
		{1-(V_m^2-|\bfM_m|^2)(N_0/2)^2+iV_mN_0}
	,
\end{align}
\end{subequations}
where the lowercase notation has been used to stress the assumed simplification.

The expansion of the $T$-matrix into charge and magnetic components further allows us to write the corrections $\delta g_0$ and $\delta\bfg_1$ to the Green function as, in general,
\begin{subequations}
\begin{align}
\delta g_0(\bfk,\bfk')=&
	g_0(\bfk)
	\sum_{mn}
		e^{-i\bfk\cdot\bfr_m}
		T_0(\bfR_{mn})
		e^{i\bfk'\cdot\bfr_n}
		g_0(\bfk')
	,
\\
\delta\bfg_1(\bfk,\bfk')=&
	g_0(\bfk)
	\sum_{mn}
		e^{-i\bfk\cdot\bfr_m}
		\bfT_1(\bfR_{mn})
		e^{i\bfk'\cdot\bfr_n}
		g_0(\bfk')
	,
\end{align}
\end{subequations}
which leads to the corresponding real space expressions
\begin{subequations}
\begin{align}
\delta g_0(\bfr,\bfr')=&
	\sum_{mn}
		g_0(\bfr-\bfr_m)
		T_0(\bfR_{mn})
		g_0(\bfr_n-\bfr')
	,
\\
\delta\bfg_1(\bfr,\bfr')=&
	\sum_{mn}
		g_0(\bfr-\bfr_m)
		\bfT_1(\bfR_{mn})
		g_0(\bfr_n-\bfr')
	.
\end{align}
\end{subequations}

With the subscripts notation $A_{pq}$ where $p=0,1$ ($q=0,1$) refers to even or odd time-reversal symmetry (parity), and we notice that
\begin{subequations}
\begin{align}
G_{00}=&
	g_0
	+
	\delta g_0
	,
	&
G_{01}=&
		0
	,
\\
\bfG_{10}=&
	\delta\bfg_1
	,
	&
\bfG_{11}=&
	0
	.
\end{align}
\end{subequations}
It can be noticed that the non-magnetic component $G_{00}$ is merely re-normalized by the presence of the magnetic defects, however, since there is no fundamental change introduced by the correction $\delta g_0$ we shall omit this contribution in the discussions below, for simplicity.
The components with $q=1$ vanish due to the absence of, for instance, spin-orbit coupling in the system. The effect of the defects is, however, to break the translation invariance in the system, something which has a profound influence on certain magnetic configurations as we shall see next.

\subsection{Double anti-ferromagnetic}
\begin{figure}[b]
\begin{center}
\includegraphics[width=0.99\columnwidth]{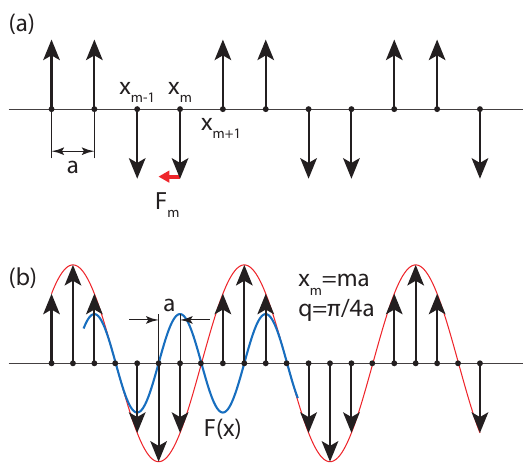}
\end{center}
\caption{(Color online) Two possible realizations of the collinear, or, sinusodal density wave. (a) Double anti-ferromagnetic structure where pairs of ferromagnetic spins are anti-ferromagnetically configured which leads to a dimerization of the ions. (b) Gradual variation of the local moment in a globally anti-ferromagnetic configuration leads to a gradual variation of the force (blue -- bold) between the ions with halved period to that of the lattice.}
\label{fig-PDW}
\end{figure}

Assume that the magnetic moments are positioned along a linear chain in $\hat{x}$-direction according to $\bfM_m\equiv\bfM(x_m)$, where $\vec{r}_m=x_m\hat{x}$ is the coordinate of the magnetic moment $\bfM_m$ and $x_{m+1}-x_m=a$. Analogously, we let $\bfQ_m\equiv\bfQ(x_m)$. We also assume the double anti-ferromagnetic structure of the magnetic moments, illustrated in Fig. \ref{fig-PDW} (a). We wish to calculate the net force exerted on the moment $\bfM_m$ by the nearest neighbor moments $\bfM_{m\pm1}$. The procedure is to evaluate the derivative $\bfF(x_m)=-(\partial/\partial\bfQ_m)\av{\Hamil_\text{MQ}}$, given the Hamiltonian
\begin{align}
\Hamil_\text{MQ}=-\frac{1}{2}
	\sum_{mn}
		\Bigl(&
			\bfM_m\cdot T^{ss}_{mn}\cdot\bfM_n
			+
			\bfM_m\cdot T^{sc}_{mn}\cdot\bfQ_n
\nonumber\\
			+&
			\bfQ_m\cdot T^{cs}_{mn}\cdot\bfM_n
			+
			\bfQ_m\cdot T^{cc}_{mn}\cdot\bfQ_n
		\Bigr)
	,
\end{align}
which gives
\begin{align}
\bfF(x_m)=
	\frac{1}{2}
	\sum_n
	\Bigl(&
		\bfM_n\cdot T^{sc}_{nm}
		+
		T^{cs}_{mn}\cdot\bfM_n
		+
		\bfQ_n\cdot T^{cc}_{nm}
		+
		T^{cc}_{mn}\cdot\bfQ_n
	\Bigr)
	.
\end{align}
In the following we shall omit the forces by the lattice-lattice coupling since we are mainly interested in the forces induced between the spin and lattice subsystems. Our interest is concerned with effects that may arise from the spin-lattice couplings $\{T^{sc}\}$ and $\{T^{cs}\}$. According to the theoretical frameworks developed in the main text, we find that we can write these interaction fields in the non-relativistic limit as
\begin{subequations}
\begin{align}
T^{sc}_{nm}=&
	-\frac{4}{\pi}
	v(x_n)\left(\im
		\int
			f(\omega)
			G_{00}(x_n,x_m)\bfG_{10}(x_m,x_n)
		d\omega\right)
	\bflambda(x_m)
	,
\\
T^{cs}_{mn}=&
	-\frac{4}{\pi}
	\bflambda(x_m)\left(\im
		\int
			f(\omega)
			\bfG_{10}(x_m,x_n)G_{00}(x_n,x_m)
		d\omega\right)
	v(x_n)
	.
\end{align}
\end{subequations}
Using the results for the Green functions derived above, we obtain, for instance,
\begin{align}
\bfG_{10}(x_m,x_n)&
	G_{00}(x_n,x_m)
\nonumber\\=&
	\delta\bfg_1(x_m,x_n)g_0(x_{nm})
\nonumber\\=&
	\sum_{\mu\nu}
		g_0(x_{m\mu})\bfT(x_{\mu\nu})g_0(x_{\nu n})
		g_0(x_{nm})
	,
\end{align}
where $x_{mn}=x_{m,n}=x_m-x_n$.

For a simple estimate of the net force we go to the limit of large separation between the defects. Then, the correction $\delta\bfg_1(x,x')=\sum_{mn}g_0(x-x_m)\bft_1(x_{mn})g_0(x_n-x')$. We also notice that $\bft_1(x_m)\sim\bfM_m$ and that $g_0(-\bfr)=g_0(\bfr)$. Considering the effects from the nearest neighbors, we then obtain
\begin{align}
\bfG_{10}(x_m,x_{m\pm1})=&
	\sum_{s=-1,0,1}g_0(x_{m,m+s})\bft_1(x_{m+s})g_0(x_{m+s,m\pm1})
\nonumber\\\sim&
	g_0(a)
	\Bigl(
		g_0(0)
		[\bfM_{m,m\pm1}+\bfM_m]
		+
		g_0(2a)\bfM_{m,m\mp1}
	\Bigr)
	,
\end{align}
where $a$ is the lattice constant ($|x_m-x_{m\pm1}|=a$). We also notice that $\bfG_{10}(x_m,x_{m\pm1})=\bfG_{10}(x_{m\pm1},x_m)$ and since $G_{00}(x_m,x_n)=G_{00}(x_n,x_m)$, it is clear that $T^{sc}_{m\pm1,m}=(T^{cs}_{m,m\pm1})^T$. Then, summarizing the force on the $m$th ion exerted by its two surrounding nearest neighbors, assuming that $v(x_m)=v$, for all $m$, under the condition that, for instance, $\bfM_{m-1}=\bfM_m=-\bfM_{m+1}$, we obtain
\begin{align}
\sum_{s=\pm1} T^{cs}_{m,m+s}\cdot\bfM_{m+s}\sim&
	-2v\bflambda(x_m)
	|\bfM_m|^2
\nonumber\\&\times
	\im
	\int
		f(\omega)
		g_0^2(a)
		\biggl(
			g_0(0)
			-
			g_0(2a)
		\biggr)
	d\omega
	.
\end{align}
Hence, the finiteness of the force on ion $m$ exerted by the nearest neighbors is determined by the real space electronic structure between the ions since $g_0(0)-g_0(2a)\sim H_0^{(1)}(0)-H_0^{(1)}(2\kappa a)\neq0$, unless $a=0$. It is therefore clear that there is a net force acting on ion $m$. The sign of the net force depends on the distance between the ions which means that the dimerization of the ions can leads to either ferromagnetic or anti-ferromagnetic pairs, details that are beyond the scope of the present context.

\subsection{Sinusodal spin density wave}
Next, we consider planar collinear, or, sinusodal spin density waves.
Therefore, we assume that the magnetic moments are positioned along a linear chain according to $\bfM_m\equiv\bfM(x_m)=M_0\hat{\bf z}\cos qx_m$, where $x_m$ is the coordinate of the magnetic moment $\bfM_m$, as is illustrated in Fig. \ref{fig-PDW} (b). Following the procedure introduced previously, we obtain the product
\begin{align}
\bfG_{10}(x_m,x_n)&
	G_{00}(x_n,x_m)
\nonumber\\=&
	\delta\bfg_1(x_m,x_n)g_0(x_{nm})
\nonumber\\=&
	\sum_{\mu\nu}
		g_0(x_{m\mu})\bfT(x_{\mu\nu})g_0(x_{\nu n})
		g_0(x_{nm})
	,
\end{align}
where $x_{mn}=x_{m,n}=x_m-x_n$.
Again, we go to the limit of large separation between the defects, which leads to that we can write
\begin{align}
\bfG_{10}(x_m,x_{m\pm1})=&
	\sum_{s=-1,0,1}g_0(x_{m,m+s})\bft_1(x_{m+s})g_0(x_{m+s,m\pm1})
\nonumber\\\sim&
	M_0g_0(a)
	\Bigl(
		g_0(0)
		[\cos qx_{m\pm1}+\cos qx_m]
\nonumber\\&
		+
		g_0(2a)\cos qx_{m\mp1}
	\Bigr)
	\hat{\bf z}
	,
\end{align}
Then, summarizing the force on the $m$th ion exerted by its two surrounding nearest neighbors, assuming that $v(x_{m\pm1})=v$, we obtain
\begin{widetext}
\begin{align}
\sum_{s=\pm1}\calT^{(cs)}_{mm+s}\cdot\bfm_{m+s}\sim&
	M_0^2
	v
	\bflambda(x_m)
	\im
	\int
		f(\omega)g_0^2(a)
		\biggl(
			\Bigl(
				g_0(0)
				[\cos qx_{m-1}+\cos qx_m]
				+
				g_0(2a)\cos qx_{m+1}
			\Bigr)
			\cos qx_{m-1}
\nonumber\\&\hspace{3.5cm}
			+
			\Bigl(
				g_0(0)
				[\cos qx_{m+1}+\cos qx_m]
				+
				g_0(2a)\cos qx_{m-1}
			\Bigr)
			\cos qx_{m+1}
		\biggr)
	d\omega
\nonumber\\=&
	M_0^2
	v
	\bflambda(x_m)
	\im
	\int
		f(\omega)g_0^2(a)
		\biggl(
			g_0(0)
			\Bigl(
				\cos^2qx_{m-1}+\cos qx_m[\cos qx_{m-1}+\cos qx_{m+1}]+\cos^2qx_{m+1}
			\Bigr)
\nonumber\\&\hspace{3.5cm}
			+
			2g_0(2a)
			\cos qx_{m-1}\cos qx_{m+1}
		\biggr)
	d\omega
	.
\end{align}
\end{widetext}
Letting $x=x_m$ such that we can write $x_{m\pm1}=x\pm a$, the trigonometric expression in the term proportional to $g_0(0)$ can be rewritten as
\begin{align}
1
	+
	\cos qa
	+
	(\cos qa+\cos2qa)\cos2qx
	,
\end{align}
whereas the corresponding expression in the term proportional to $2g_0(2a)$ as
\begin{align}
\frac{1}{2}\Bigl(
	\cos2qx+\cos2qa
	\Bigr)
	.
\end{align}
With these equalities, we can write the force as proportional to
\begin{align}
M_0^2v\bflambda(x)&
	\im
	\int
		f(\omega)g_0^2(a)
		\biggl(
			g_0(0)[1+\cos qa]+g_0(2a)\cos2qa
\nonumber\\&
		+
			\Bigl(
				g_0(0)[\cos qa+\cos2qa]
				+
				g_0(2a)
			\Bigr)
			\cos2qx
		\biggr)
	d\omega
	.
\end{align}
Here, taking $q=\pi/4a$, see Fig. \ref{fig-PDW} (b), this expression reduces to
\begin{align}
\frac{\sqrt{2}}{2}M_0^2v\bflambda(x)&
	\im
	\int
		f(\omega)
		g_0^2(0)
		\biggl[
			\Bigl(
				1+\sqrt{2}
			\Bigr)
			g_0(0)
\nonumber\\&
			+
			\Bigl(
				g_0(0)+\sqrt{2}g_0(2a)
			\Bigr)
			\cos\frac{\pi x}{2a}
		\biggr]
	d\omega
	.
\end{align}
The spatial variation of the resulting forces has a period which is half of that of the lattice.

\bibliographystyle{apsrev4-1}
\bibliography{short,refs,librarySLD}

\end{document}